\documentclass[transmag]{IEEEtran}
\usepackage{latexsym}
\usepackage{graphicx}
\usepackage{amsfonts,amssymb,amsmath,comment}
\usepackage{hyperref}
\usepackage{bm}
\usepackage{mathtools}
\usepackage{xparse}
\usepackage{graphicx}               
\usepackage{amsfonts}
\usepackage[bbgreekl]{mathbbol}
\usepackage{setspace}
\usepackage{algorithm,algpseudocode}
\usepackage{amsmath}
\usepackage{amsfonts}
\usepackage{amssymb}

\algnewcommand{\IIf}[1]{\State\algorithmicif\ #1}
\algnewcommand{\EndIIf}{\unskip\ \algorithmicend\ \algorithmicif}
\usepackage{verbatim}
\usepackage{blindtext}
\usepackage{xcolor}

\usepackage{amsmath}
\usepackage{amssymb}
\usepackage{subcaption}
\usepackage{lipsum}
\usepackage{hyperref}
\usepackage{setspace}
\renewcommand{\arraystretch}{0.6}

\usepackage{mathrsfs}
\usepackage{caption}
\usepackage{diagbox}

\usepackage{arydshln}
\usepackage{multicol}
\usepackage{verbatim}
\usepackage{tikz}
\usepackage{cite}
\usepackage{setspace} 

\usepackage{xcolor}
\title{R-local unlabeled sensing: A novel graph matching approach for multiview unlabeled sensing under local permutations}

\author{Ahmed Ali Abbasi, Abiy Tasissa,  Shuchin Aeron
\thanks{The authors are with Tufts University, Medford, USA e-mail:ahmed.abbasi@tufts.edu}
\thanks{This research is supported by NSF CCF:1553075, NSF TRIPODS grant HDR:1934553, and AFOSR FA9550-18-1-0465.r}}

\newtheorem{definition}{Definition}

\def\A{\mathbf{A}}
\def\B{\mathbf{B}}
\def\C{\mathbf{C}}
\def\D{\mathbb{D}}
\def\Y{\mathbf{Y}}
\def\Z{\mathbf{Z}}
\def\P{\mathbf{P}}
\def\X{\mathbf{X}}
\def\pi{\mathbb{\Pi}}
\def\real{\mathbb{R}}
\def\tr{\text{trace}}
\begin{document}

\IEEEtitleabstractindextext{\begin{abstract}
Unlabeled sensing is a linear inverse problem where the measurements are scrambled under an unknown permutation leading to loss of correspondence between the measurements and the rows of the sensing matrix. Motivated by practical tasks such as mobile sensor networks, target tracking and the pose and correspondence estimation between point clouds, we study a special case of this problem restricting the class of permutations to be local and allowing for multiple views. In this setting, namely multi-view unlabeled sensing under local permutations, previous results and algorithms are not directly applicable. In this paper, we propose a computationally efficient algorithm, R-local unlabeled sensing (RLUS), that creatively exploits the machinery of indefinite relaxations of the graph matching problem  to estimate the local permutations. Simulation results on synthetic data sets indicate that the proposed algorithm is scalable and applicable to the challenging regimes of low to moderate SNR. \end{abstract}

}

\maketitle
\section{Introduction}
\label{sec:intro}

Motivated by several practical problems such as sampling in the presence of clock jitter, mobile sensor networks and multiple target tracking in radar, the problem of unlabeled sensing was first considered in \cite{unnikrishnan2018unlabeled}. There, the authors derived information theoretic results for identification of unknown signal under linear measurements when the measurement correspondence is lost. Several generalizations of this problem as well as specific cases have been considered in \cite{abid2018stochastic,homomorphic_sensing,AI-EM,pananjady,slawski-single,hsu2017linear,compressed,LEVSORT,rahmani,spherical,zhang2019permutation,slawski_two_stage}. A detailed survey of these works vis-a-vis our work is outlined in the related work section \ref{subsec:relation}. Below we first introduce and motivate the model considered here. 

\begin{figure}[htbp]
    \centering
    {\includegraphics[width=0.40\linewidth,height = 3.0 cm]{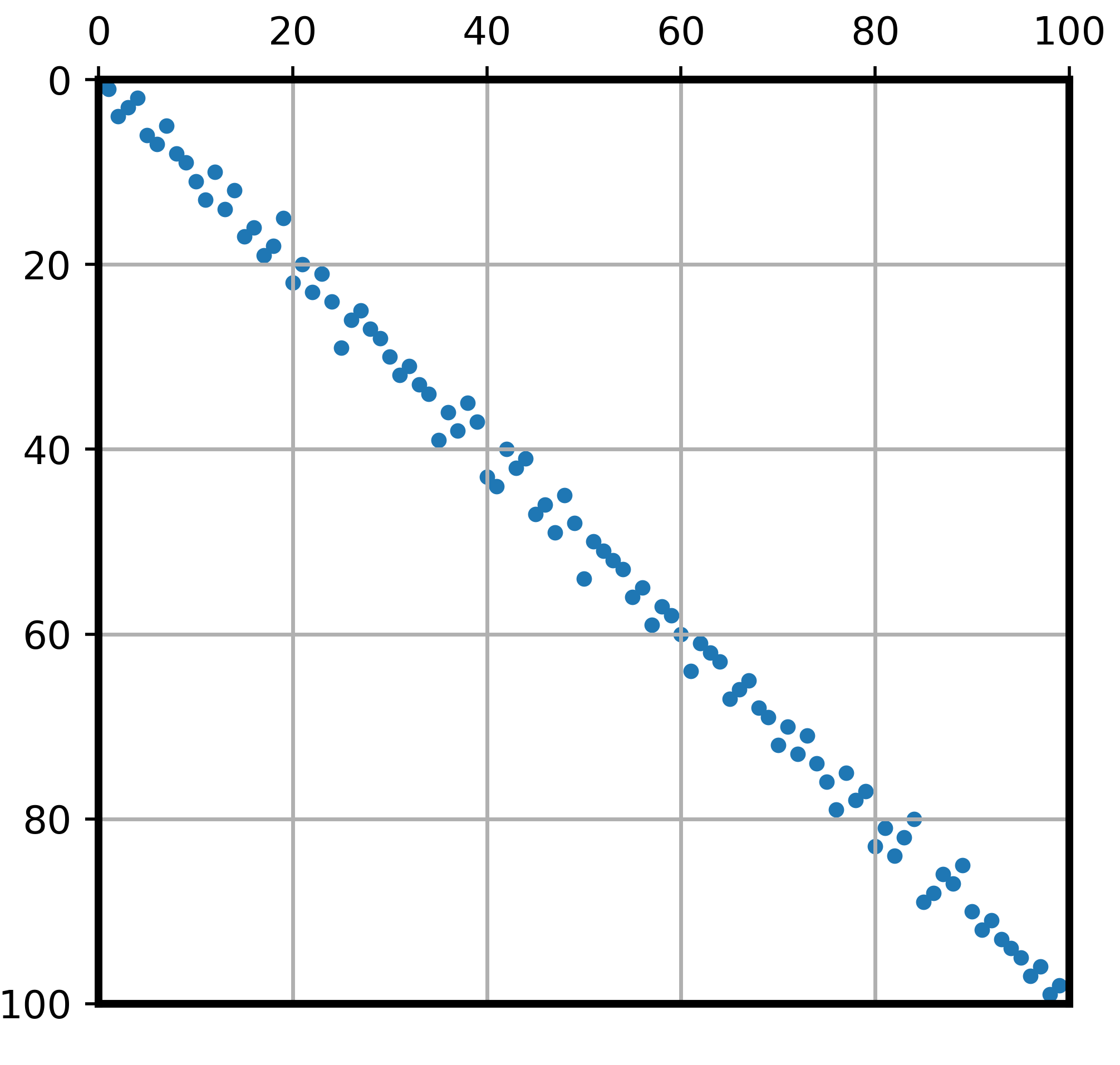}}
    {\includegraphics[width=0.40\linewidth,height = 3.0 cm]{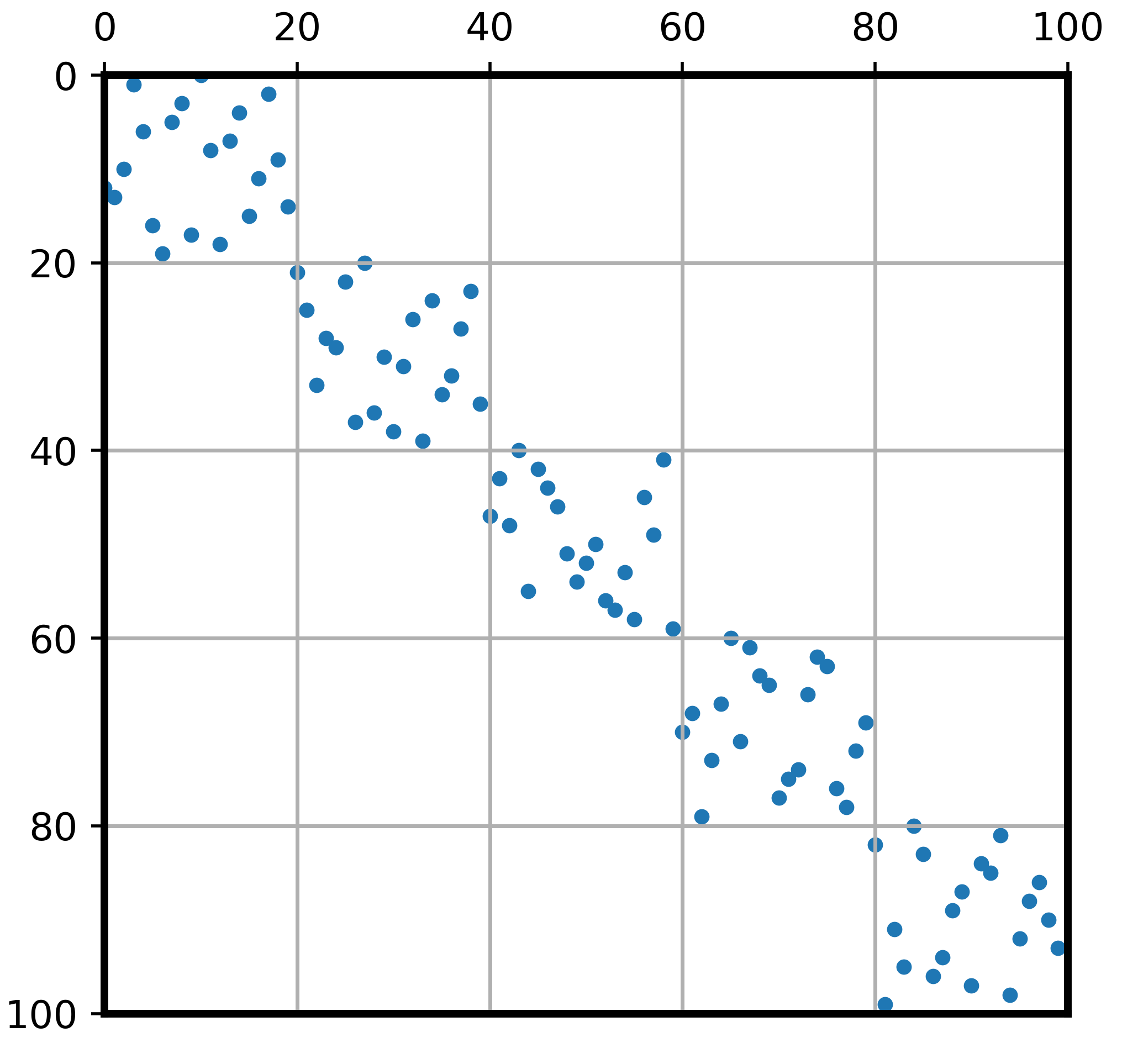}}
    \caption{Left: The sparse permutation with $r$-local structure considered in \cite{spherical} with $r$ = 20. A permutation is sparse \cite{rahmani} if it has a small number of off diagonal elements. Right: $20$-local permutation without sparsity.}
    \label{fig:Local-Permutations}
\end{figure}

In this paper, we study a specific case of the unlabeled sensing problem where the set of permutations, modeling the loss of correspondence, is local. Specifically, we consider  the task of estimating the unknown permutation $\mathbf{\P}^* \in \pi_{n,r}$ from the views $\Y \in \real^{n \times m}$ such that 
\begin{equation}
    \Y = \boldsymbol{\P}^{*} \B \X^* + \mathbf{N}.
    \label{eq:local}
\end{equation}
In the representation above, $\B \in \real^{n \times d}$ is the known measurement matrix, $\mathbf{N} \in \real ^ {n \times m}$ is additive noise with i.i.d entries from the Gaussian distribution $\mathcal{N}\left(0,\sigma^{2}\right)$ and $\P^{*}\in \pi_{n.r}$ is an $r$-local permutation  defined below.
\begin{definition}
Permutation ${\boldsymbol{\P}}$ is ${r}$-local, \textit{i.e.} $\P \in \pi_{n,r}$, if it is composed of ${n/r}$ blocks along the diagonal, with each block being a permutation matrix of size $r \times r$. More precisely, $\mathbf{\P} = \textrm{diag}\left(\P_1, \cdots, \P_{n/r}\right)$, where $\P_{i} \in \pi_r$ denotes an $r \times r$ permutation matrix. 
\label{def1}
\end{definition}
Fig. \ref{fig:Local-Permutations} shows examples of $r$-local permutations for $r = 20$.

\subsection{Motivating Applications}
We outline two applications that can fall under the proposed model. Consider samples ${y}(i) = {x}(iT)$  of an analog signal $\bm{x}(t)$. Jittered sampling \cite{Balakrishnan}, defined as ${y}(i) = {x}(iT + T_{\textrm{skew}})$, in interleaved ADC systems \cite{jitter} scrambles the order of the $r$ samples output at each cycle.

\begin{figure}[t!]
    \centering
    {\includegraphics[width = 0.75\linewidth ]{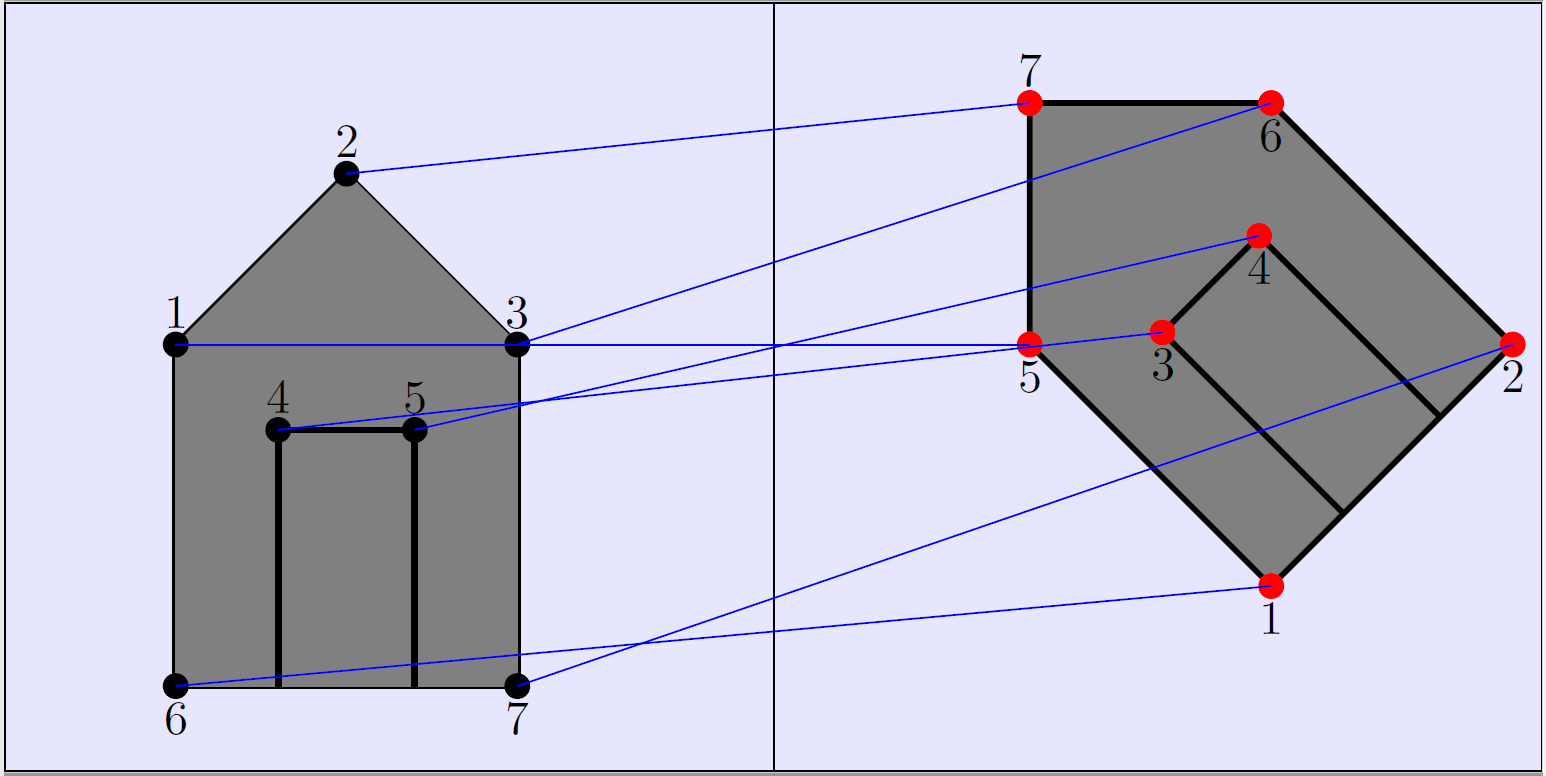}}
        \caption{Left: An object with $7$ keypoints, Right: The object after linear transformation (a rotation of 45 degrees) with 7 keypoints. In the pose and correspondence problem, given the keypoints corresponding to the left  and right images, the task is to find the underlying permutation.}
        \label{fig:pose_and_correspondence_illustration}
\end{figure}

The second application is the pose and correspondence problem and is illustrated in Fig. \ref{fig:pose_and_correspondence_illustration}.

\label{subsec:apps}
\subsection{Relation to Existing Work}
\label{subsec:relation}
\begin{table*}
    \renewcommand*{\arraystretch}{1.1}
    \centering
    \begin{tabular}{|l|l|l|}
        \hline
       \textbf{ Paper} & \textbf{Model} $\left( \P^{*} \mid \mathbf{X}^{*}_{d \times m} \mid \sigma^{2} \right)$ & \textbf{Approach} \\
        \hline
        \cite{abid2018stochastic} & $\P^{*} \in \mathbb{\Pi}_n \mid m = 1 \mid  \sigma^{2} > 0$ & Likelihood maximization\\
        \hline 
        \cite{homomorphic_sensing} & $\P^{*} \in \mathbb{\Pi}_n \mid m = 1 \mid  \sigma^{2} > 0$ & Dynamic programming\\
        \hline 
        \cite{AI-EM} & $\P^{*} \in \mathbb{\Pi}_n \mid m = 1 \mid  \sigma^{2} > 0$  & Symmetric Polynomials\\
        \hline
        \cite{pananjady} & $\P^{*} \in \mathbb{\Pi}_n \mid d = 1 , m = 1 \mid \sigma^{2} > 0$ & One dimensional sort\\
        \hline
        \cite{slawski-single} & $\P^{*} \in \mathbb{\Pi}_{n,s} \mid m = 1 \mid  \sigma^{2} > 0$ & Linear program for $\P^{*}$, convex program for $\mathbf{X}^{*}$\\
        \hline
        \cite{hsu2017linear} & $\P^{*} \in \mathbb{\Pi}_n\mid m = 1, \text{ fixed } \bm{x}^* \mid  \sigma^{2} = 0$ & Lattice reduction\\
        \hline
        \cite{compressed} & $\P^{*} \in \mathbb{\Pi}_n \mid m = 1, \text{ sparse } \bm{x}^* \mid  \sigma^{2} = 0$  & Branch and bound\\
        \hline
        \cite{LEVSORT} & $\P^{*} \in \mathbb{\Pi}_n \mid m = d \mid  \sigma^{2} = 0$  & Subspace matching \\
        \hline
        \cite{rahmani} &  $\P^{*} \in \mathbb{\Pi}_{n,s} \mid m = d \mid  \sigma^{2} > 0$  &  Subspace clustering\\
        \hline 
        \cite{spherical} & $\P^{*} \in \mathbb{\Pi}_{n,s}  \mid m = d, \mathbf{X}^{*}(\X^{*})^{\intercal} = \mathbf{I}  \mid 
        \sigma^{2} > 0$ & Orthogonal least squares \\
        \hline
        \cite{zhang2019permutation}  & $\P^{*} \in \mathbb{\Pi}_n \mid m > 1 \mid  \sigma^{2} > 0$ & Linear program for $\P^{*}$, closed form solution for $\mathbf{X}^{*}$ \\
        \hline 
        \cite{slawski_two_stage} & $\P^{*} \in \mathbb{\Pi}_{n,s} \mid m > 1 \mid  \sigma^{2} > 0$ & Linear program for $\P^{*}$, convex program for $\mathbf{X}^{*}$ \\
        \hline
        \textbf{This paper} &  $\P^{*} \in \pi_{n,r} \mid m > 1 \mid  \sigma^{2} > 0$ & Graph matching via quadratic assignment problem\\
        \hline
    \end{tabular}
\caption{Summary of Existing Works. $\mathbb{\Pi}_{n}$ denotes the general set of permutations of order $n$. $\mathbb{\Pi}_{n,s}$ denotes the set of sparse permutations. This paper considers the set of $r$-local permutations, $\pi_{n,r}$, see \textit{Definition} 1, Fig. \ref{fig:Local-Permutations}. }
\label{Table:relation}
\end{table*}
The problem of unlabeled sensing has received considerable attention. In addition to the fully shuffled case, several variants with structure on either $\X^*$ and/or the underlying permutation $\mathbf{\P}^{*}$ have been studied. The single view case is a specific instance of the problem in \eqref{eq:local}, with $m=1$, where the view matrix $\Y$ reduces to a single column vector.  In TABLE \ref{Table:relation}, we summarize the various models, based on different structures on $\X^{*}$ and $\mathbf{\P}^{*}$, and the corresponding algorithms. The rest of the literature survey is focused on the multi-view unlabeled sensing problem, the subject of the present paper.

The LEVSORT algorithm in \cite{LEVSORT} recovers the unknown permutation by aligning the leverage scores corresponding to the measurements matrix $\Y$ and sensing matrix $\B$. A drawback of this algorithm is its sensitivity to additive noise. In \cite{rahmani}, another structure on $\mathbf{\P}^{*}$ is considered for the multiview setting, namely that the number of shuffled elements is much smaller compared to $n$.  By observing that the inliers lie in the span of a lower $d$-dimensional subspace of $\real^{n}$, the authors in \cite{rahmani} consider a sparse subspace clustering approach. For the same model proposed in \cite{rahmani}, namely sparse $\mathbf{\P}^{*}$, the works in \cite{slawski-single,slawski_two_stage} proposed $\ell_{1}$ constrained robust regression formulations.

In addition to assuming structure on the unknown permutation $\P^{*}$, previous techniques are applicable when dimension $d$ is small \cite{AI-EM,homomorphic_sensing}. The results in \cite{LEVSORT,rahmani,spherical} require number of views $m$ equal to dimension $d$.  By contrast, we consider recovery under a more challenging regime where $n$ scales moderately with $d$, a small number of views with $m<d$ and additive measurement noise.  Finally, the block diagonal permutation structure we consider has also been discussed in \cite{spherical} but with the following additional assumptions on the model in \eqref{eq:local}: (a) The data matrix $\X^* \in \real^{d \times d}$ is orthonormal, and (b) The block diagonal permutation $\boldsymbol{\P}^{*}$ is sparse. The discussion in section \ref{subsec:proposed_approach} shows that the assumptions in \cite{spherical} result in a simpler instance of the graph alignment problem compared to the problem considered in this paper. A related problem in graph signal processing (GSP) is the recovery of a band-limited graph signal \cite{lorenzo2018sampling,tsitsvero2016signals}. However, in the setup we consider here, there is no feature associated with nodes that will help us in  graph matching. Since we are agnostic to what the nodes represent and the network topology, the GSP framework is not applicable here. 

\paragraph{Major Contributions} Our main contributions can be summarized as follows. 
\begin{enumerate}
\item For the unlabeled sensing problem, we present a novel graph matching formulation and propose an algorithm, R-local unlabeled sensing (RLUS), that creatively exploits the machinery of graph matching to resolve local permutations via multiple views. 
\item We propose an initialization that utilizes the r-local structure based on collapsing the measurements. This principled initialization shows competitive performance in numerical simulations. 
\item Motivated by several applications, we analyze a novel permutation model that we refer to as the $r$-local model (see Def. \ref{def1}). A very restrictive version of the $r$-local model was considered in \cite{spherical} and the setup that we consider here significantly expands upon the model therein.
\item One of the main conclusions of this paper is that, for the $r$-local model, multiple views significantly help in permutation recovery.
\end{enumerate}
\paragraph{Outline}
The rest of the paper is organized as follows: Section \ref{subsec:proposed_approach} gives background on graph matching and outlines our proposed approach. In section \ref{sec:algo}, our proposed algorithm is discussed. In section \ref{sec:results}, detailed simulation results are provided. We conclude in Section \ref{sec:conclude} with the main findings of the paper.
\subsection{Notation}
\label{subsec:notation}
The following is a summary of the notation used in this paper. Let $\bm{b}_{i}^{\intercal} \hspace{0.1cm} \forall \hspace{0.05cm} i \hspace{0.05cm} \in [n]$ denote the rows (measurement vectors) of the matrix $\B \in \real^{n \times d}$ i.e. $\B^{\intercal} = \left[\bm{b}_{1}^{\intercal} \mid,\cdots,\mid \bm{b}_{n}^{\intercal}\right]$. Similarly $\Y^{\intercal} = \left[\bm{y}_{1}^{\intercal} \mid,\cdots,\mid \bm{y}_{n}^{\intercal}\right]$. Let $\Y_{k} \in \real^{r \times d} \hspace{0.1cm} \forall \hspace{0.05cm} k \hspace{0.05cm} \in [n/r]$ denote blocks of $\Y$ i.e. $\Y^{\intercal} = [\Y_{1}^{\intercal} \mid \cdots \mid \Y_{n/r}^{\intercal} ]$. Let $\Y_{P_k,:}$ denote the submatrix of $\Y_k$  comprising rows with indices in $P_k$. $\mathbb{1}_{r} \in \real^{r}$ denotes the $r$-dimensional vector of all ones. $\pi_n$ denotes the set of permutations in $\real^{n \times n}$. $\D_n \in \real^{n \times n}$ is the set of doubly stochastic matrices: $ \D_n = \{\P \in \real^{n \times n} \mid \P\mathbb{1}_n = 1, \P^{T}\mathbb{1}_n = 1, P_{ij} \geq 0 \}$. $\mathbf{I}$ denotes the identity matrix. The Moore-Penrose pseudoinverse of a matrix is denoted by $(\cdot)^{\dagger}$. $\lVert \cdot \rVert_F^2$ denotes the square of the Frobenius matrix norm defined as the sum of the squares of the elements of the matrix.
\section{Proposed approach}
\label{subsec:proposed_approach}

A central part of the proposed algorithm is based on theory and methods for graph matching, which we briefly discuss here. Given two graphs $G_1 = (V_1,E_1)$ and $G_2 = (V_2,E_2)$ each with $n$ vertices, the graph matching problem is concerned with finding the permutation or map that best aligns them. If we choose the Frobenius norm as a metric of alignment, the general graph matching problem can be formulated as follows
\begin{equation}\label{eq:gm}
\underset{\P \in \pi_n }{\min}\quad ||\A\P-\P\B||_{F}^{2},
\end{equation}
where $\A$ and $\B$ are weighted adjacency matrices of $G_1$ and $G_2$ respectively. Using the fact that $\P$ is an orthonormal matrix, the objective can be expanded as follows: $||\A\P-\P\B||_{F}^{2} = ||\A||_F^2+||\B||_F^2-2\,\tr(\A\P\B^{\intercal}\P^{\intercal})$. With this, \eqref{eq:gm}
can also be equivalently written as
\begin{equation}\label{eq:gm_qap}
\underset{\P \in \pi_n }{\min}\quad -\tr(\A\P\B^{\intercal}\P^{\intercal})
\end{equation}
This latter problem is the well known quadratic assignment problem (QAP) \cite{graph_matching}. 

Since the minimization problem over the set of permutations is NP hard, a common relaxation of the graph matching programs in \eqref{eq:gm} and \eqref{eq:gm_qap} is based on replacing the set of permutation matrices with the set of doubly stochastic matrices $\D_n$. Using the Birkhoff-Von Neumann theorem \cite{birkhoff1946three,von1953certain}, $\D_n$ is known to be the convex hull for the set of permutation matrices. While the objectives in \eqref{eq:gm} and \eqref{eq:gm_qap} are equivalent over the set of permutation matrices, they are markedly different after relaxation. The relaxation of \eqref{eq:gm} is a convex program while the relaxation of \eqref{eq:gm_qap} is typically indefinite and generally a nonconvex program. Here on, we refer to this relaxation as the indefinite relaxation. The theoretical analysis on correlated random Bernouilli graphs in  \cite{lyzinski2015graph}
shows that the indefinite relaxation recovers the true permutation of exact graph matching problem with high probability while the convex program almost always fails. In addition, the indefinite relaxation is shown to be empirically successful for fast graph matching \cite{fast_gm}. This relaxation is utilized in this paper. 
\subsection{Covariance matching: Orthonormal $\X$}
\label{subsec:orth_X}
Given permuted observations $\Y$, the sensing matrix $\B$ and with $\C_Y$ and $\C_{\widehat Y}$ denoting covariance matrices, the unknown permutation is estimated as follows
\begin{equation}
    \widehat{\P} = \underset{{\widehat{{\Y}} \in \real^{n \times m} , \P \in \pi_n}}{\arg \min} {\lVert  \C_{Y} \P - \P\C_{\widehat{{Y}}} \rVert}^{2}_{F},
    \label{eq:approach}
\end{equation}
where $\widehat{\Y} = \B{\widehat{\X}}$, $\C_{\widehat  Y} = \widehat{\Y}{\widehat{\Y}}^{\intercal}$, ${\C_{Y}} = \Y\Y^{\intercal}$ and $\widehat{\P}$ is an estimate of $\P^{*}$. For a fixed $\widehat{\Y}$, the minimization problem in \eqref{eq:approach} corresponds to a QAP with similarity matrices $\C_Y$ and $\C_{\widehat Y}$. However, unlike \eqref{eq:gm_qap} where the weighted adjacency matrices are explicitly given, $\C_{\widehat Y}$ in \eqref{eq:approach} has to be estimated as $\X$ is unknown. When $\X$ is an orthonormal matrix, as considered in \cite{spherical}, $\C_{\widehat Y} = \B\B^{\intercal}$ does not depend on $\X$. In this case, it is easy to see that the  alignment in \eqref{eq:approach} reduces to:
\begin{equation}
    \widehat{\P} = \underset{ \P \in \pi_n}{\arg \min}\quad  {\lVert \C_{Y} \P - \P\C_{B} \rVert}^{2}_{F}, \label{orth_X}
\end{equation}
with $\C_{B}= \B\B^{\intercal}$.
\begin{figure}[t!]
\begin{subfigure}{.15\textwidth}
        \begin{center}
            {\includegraphics[width = \linewidth ]{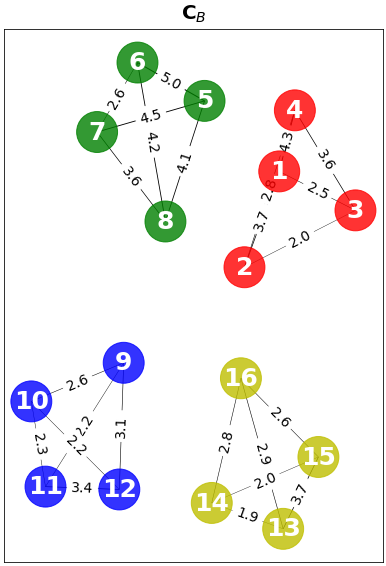}}
        \end{center}
\end{subfigure}
\begin{subfigure}{.15\textwidth}
        \begin{center}
            {\includegraphics[width = \linewidth]{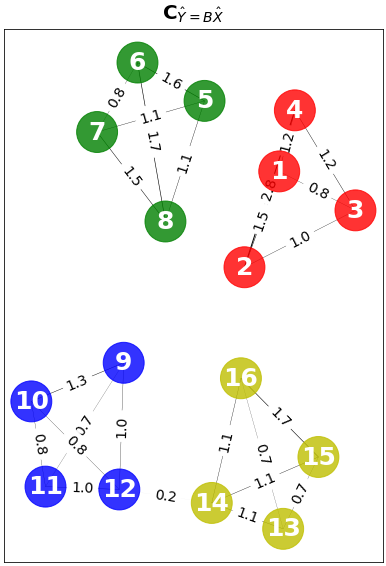}}
        \end{center}\end{subfigure}
\begin{subfigure}{.15\textwidth}
        \begin{center}
            {\includegraphics[width = \linewidth]{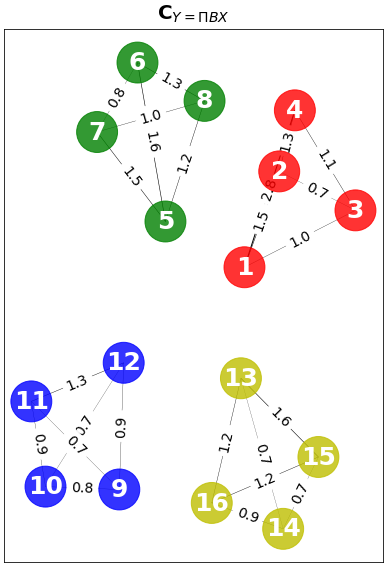}}
        \end{center}\end{subfigure}
\caption{A block of $r$ consecutive rows in matrices $\B,\Y,\widehat{\Y}$ corresponds to a cluster of $r$ nodes in the corresponding graph. Left:  The graph of known measurement matrix $\B$. Middle: The graph of given  measurements $\Y = \P^* \B\X$. The $r$-local permutation, $\P^*$ distorts correspondence of nodes within each cluster, and the unknown linear transformation $\X$ distorts the edge weights. Right: The graph of estimate $\widehat{\Y} = \B\widehat{\X}$. The proposed approach first estimates $\widehat{\Y}$ and estimates $\mathbf{\P}^*$ by matching $\C_{ \widehat Y}$ to $\C_Y$. }
\label{fig:proposed_approach}
\captionsetup{belowskip=0pt}
\end{figure}
\subsection{Covariance matching: General $\X$}
For general $\X$, we could consider an alternating minimization between $\widehat{\mathbf{\P}},\widehat{\X}$. Since minimizing \eqref{eq:approach} is an NP-hard problem \cite{graph_matching}, we propose a two-stage algorithm to first estimate $\widehat{\X}$. We set $\widehat{\Y} = \B\widehat{\X}$ and let $\widehat{\mathbf{\P}}$ be given via the indefinite relaxation of \eqref{eq:approach}.

\noindent \textbf{Stage-A}: In stage-A of the algorithm, we employ alternating minimization over $\X,\pi_n$on the following bi-convex objective
\begin{equation}
    (\widehat{\X},\widehat{\mathbf{\P}}) = \underset{\X,\P\in \pi_n}{\arg \min}\quad \lVert \Y - \mathbf{\P} \B \X\rVert_F^2
    \label{eq:LP}
\end{equation}
We leverage the $r$-local assumption on $\mathbf{\P}^{*}$ for initializing $\widehat{\X}$ for the optimization in \eqref{eq:LP}. This will be done via the collapsed linear system in  \eqref{collapsed} to be discussed later. Unlike the QAP in \eqref{eq:approach}, the optimization over $\pi_n$ in \eqref{eq:LP} is a linear programming problem over $\D_n$. As outlined in section \ref{sec:algo}, the $r$-local assumption on $\mathbf{\P}^*$ allows us to minimize over $\pi_n$ locally by using permutations of size $r\times r$.
\begin{align*}
\widehat{\P}_k & = \underset{\P_k \in \pi_r}{\arg \min} \quad  \lVert \Y_k - \mathbf{\P}_k \B_k {\widehat{\X}}\rVert_F^2 \\
& = \underset{\P_k \in \pi_r}{\arg \min}-\tr (\B_k\widehat{\X}\Y_k^{\intercal} \mathbf{\P}_k),
\end{align*}
$\forall  k \in [n/r]$ with $\B_k$ and $\Y_k$ denoting the blocks of matrices $\B$ and $\Y$ respectively \footnote{Section \ref{subsec:notation}}. 

\noindent \textbf{Stage-B}: Given estimate $\widehat\Y = \B \widehat \X$ from stage-A, we employ the indefinite relaxation of graph matching to estimate $\widehat{\P}$. This will be done via estimating the blocks of $\widehat{\P}$ as follows.
\begin{equation}
    \widehat{\P}_k =  \underset{\P_k \in \D_{r}} {\arg \min} \quad  -\tr (\C_{k} \P_k \widehat{\C}_{k}^{\intercal} \P_k^{\intercal}),
    \label{eq:rQAP}
\end{equation}
$\forall  k \in [n/r]$ with ${\widehat{\C}}_{k} = \widehat{\Y}_k \widehat{\Y}_k^{\intercal}$ and $\C_{k} = \Y_k \Y_k ^{\intercal}$ denoting the block covariance matrices.

Fig. \ref{fig:proposed_approach} is a visual summary of the proposed approach. 
With this perspective, in section \ref{sec:algo}, we propose and outline the algorithm in detail for the $r$-local permutation model. In section \ref{sec:results}, we provide simulation results, and focus particularly on quantifying permutation recovery via the proposed algorithm under an increasing number of views.
\section{R-local unlabeled sensing (RLUS) algorithm}
RLUS has three main components: (a) an initialization via collapsing, (b) estimating $\widehat{\Y}$ and (c) estimating $\widehat{\P}$. This will be discussed in detail below.
\subsection{Initialization}
\label{sec:algo}
\begin{algorithm}[t]
\caption{Stage-A}
\label{stage_A}
\begin{algorithmic}[1]
\Require{$\text{Radius }r \in \real, \B \in \real^{n \times d} , \Y \in \real^{n \times m}$}
\Ensure{$\widehat{\Y}, \widehat{\P}_{\mathrm{LP}}$}
\State{$P \gets \{1, \cdot \cdot , n\} \enskip Q \gets \{1, \cdot \cdot, n\} \enskip K^{(0)} \gets \{1, \cdot \cdot,  n/r\}$} \vspace{0.1em}
\State{$\tilde{\bm{b}}^{{\intercal}^{(0)}}_{i} \gets \mathbb{1}_{i}^{\intercal}\B \quad  \quad \tilde{\bm{y}}_{i}^{\intercal^{(0)}} \gets \mathbb{1}_{i}^{\intercal}\Y \hspace{0.35cm} \forall  i \in [{n}/{r}]$}\vspace{0.15em} 
\State{$\widehat{\Y}^{(0)} \gets \B(\widetilde{\B}^{\dagger}\widetilde{\Y}) $}
\For {$t \in 0 \cdots d - n/r - 1$}
    \For {$k \in K^{(t)}$}
    \State{$k_r \gets \{(k-1)r+1,\cdots,kr\}$}
    \State{$P_k \gets k_r \cap P \quad Q_k \gets k_r \cap Q $  }
    \State{$\widehat{\P}_k = \underset{\P \in \pi_{\lvert P_k \rvert}}{\arg \min} -\tr{(\widehat{\Y}^{(t)}_{P_k,:}\Y^{\intercal}_{Q_k,:}\P)} $} 
    \State{$P_k \gets \widehat{\P}_k P_k$}\vspace{0.15cm}
    \EndFor
    \State{$(p^{*(t)},q^{*(t)}) = \underset{p \in P_k,q \in Q_k}{\arg \min} {{\big \lVert \Y - \B \big( \big[ \frac{\widetilde{\B}^{(t)}}{b^{\intercal}_p} \big]}^{\dagger} \big[\frac{ \widetilde{\Y}^{(t)}}{y^{\intercal}_q} \big]\big) \big \rVert}_{\mathrm{F}} $} \vspace{0.11cm}
    \State{$\widetilde{\B}^{(t+1)} \gets \big[ \frac{\widetilde{\B}^{(t)}}{b^{\intercal}_{p^{*}}} \big] \quad \widetilde{\Y}^{(t+1)} \gets \big[ \frac{\widetilde{\Y}^{(t)}}{y^{\intercal}_{q^{*}}} \big]$} \vspace{0.11cm}
    \State{$\widehat{\Y}^{(t+1)} \gets \B (\widetilde{\B}^{(t+1)\dagger}\widetilde{\Y}^{(t+1)})$} \vspace{0.11cm}
    \State{$P \gets P \setminus p^* \quad Q \gets Q \setminus q^*$} \vspace{0.5em}
    \For {$k \in K^{(t)}$}
    \State{$k_r \gets \{(k-1)r+1,\cdots,kr\}$}
    \IIf{$( \lvert k_{r} \cap P \rvert = 1) \enskip K^{(t+1)} \gets K^{(t)} \setminus k $}
    \EndFor
\EndFor
\For {$k \in 1, \cdots, n/r$}
\State{$\widehat{\P}_k = \underset{\P \in \pi_r}{\arg \min} -\tr{(\widehat{\Y}^{(t)}_{k}{\Y}^{\intercal}_{k}\P_k)}$}
\EndFor
\State{$\widehat{\mathbf{\P}}_{\text{LP}} \gets \text{diag} (\widehat{\P}_{1},\cdots,\widehat{\P}_{n/r})$} \vspace{0.11cm}
\State{$\widehat{\Y} \gets \B(\B^{\dagger}\widehat{\mathbf{\P}}_{\text{LP}}^{\intercal}\Y)$} \vspace{0.11cm}
\end{algorithmic}
\end{algorithm} 
\begin{algorithm}
\caption{Stage-B}
\label{stage_B}
\begin{algorithmic}[1]
\Require{$\text{Radius }r , \B  , \Y ,\widehat{\Y} , \widehat{\P}_{\mathrm{LP}}= \textrm{diag}(\bar{\P}_1,\cdots,\bar{\P}_{n/r})$}
\Ensure{$\widehat{\mathbf{\P}}_r$}\vspace{0.1em}
\For {$k \in 1 \cdots n/r$}\vspace{0.15em}
\State{$\widehat{\C}_{k} \gets \widehat{\Y}_{k}\widehat{\Y}_{k}^{\intercal} \quad {\C}_{k}  \gets  {\Y_{k}\Y_{k}^{\intercal}}    $}\vspace{0.11cm}
\State{$\widehat{\P}_k = \underset{\P_k \in \D_r}{\arg \min} -\tr ({\C}_{k} \P_k \widehat{\C}_{k}^{\intercal}  \P_k^{\intercal})$ //\textit{Initialize to $\bar{\P}_{k}$}} 

\State{$\widehat{\P}_k = \underset{\P_k \in \pi_r}{\arg \min} -\tr{(\widehat{\P}^{\intercal}\P_k)}$}
\EndFor
\State{$\widehat{\mathbf{\P}} \gets  \text{diag} (\widehat{\P}_{1},\cdots,\widehat{\P}_{n/r})$} 
\end{algorithmic}
\label{algo:gm}
\end{algorithm} 
For our problem setup, we are given a sensing matrix $\B$, locally permuted observations $\Y$ and radius $r$. We note that, for $\X \in \real^{d \times m}$,  each $r \times d$ block of measurement vectors and $r \times m$ block of permuted observations, $\left[
\begin{array}{c:c}
{\P_k\B_k} & {\Y_k}
\end{array}\right] \enskip \forall \enskip k \in [n/r]$, is an $r$-dimensional unlabeled sensing problem. We can extract $n/r$ labeled measurements on $\X$ by collapsing each block as follows
\begin{align*}
    \P_{1}\B_{1} \X & = \Y_{1} & \implies \hspace{1cm} \underbrace{\mathbb{1}_{r}^{\intercal}\P_{1}\B_{1}}_{\tilde{\bm{b}}_{1}^{\intercal}} \X & = \underbrace{\mathbb{1}_{r}^{\intercal}\Y_{1}}_{\tilde{\bm{y}}_{1}^\intercal}\\ 
    \P_{2}\B_{2} \X & = \Y_{2} & \implies \hspace{1cm} \underbrace{\mathbb{1}_{r}^{\intercal}\P_{2}\B_{2}}_{\tilde{\bm{b}}^{\intercal}_{2}} \X & = \underbrace{\mathbb{1}_{r}^{\intercal}\Y_{2}}_{\tilde{\bm{y}}_{2}^{\intercal}} \\
    & \vdots &  & \vdots\\                                      
    \P_{{n}/{r}}\B_{{n}/{r}} \X & = \Y_{n/r} & \implies \hspace{0.45cm} \underbrace{{\mathbb{1}_{r}^{\intercal}}\P_{{n}/{r}}\B_{{n}/{r}}}_{\tilde{\bm{b}}^{\intercal}_{n/r}}\X & = \underbrace{\mathbb{1}_{r}^{\intercal}\Y_{n/r}}_{\tilde{\bm{y}}_{n/r}^{\intercal}}
\end{align*}
 to form the \textit{collapsed} system
 \begin{equation}
\left[
\begin{array}{c:c}
\widetilde{\B} & \widetilde{\Y}
\end{array}
\right]
\in \real^{({n}/{r}) \times m}
\label{collapsed}
\end{equation}
While the permuted linear system of equations (L.S.E.)
$
\left[
\begin{array}{c:c}
{\P^*\B} & {\Y}
\end{array}\right]
\in \real^{n}$ has $n>d$ permuted measurements, the collapsed L.S.E.
$\left[
\begin{array}{c:c}
\widetilde{\B} & \widetilde{\Y}
\end{array}
\right]
\in \real^{{n}/{r}\times m}$ is underdetermined with ${{n}/{r}}$ measurements. We consider recovery of $\mathbf{\P}^*$ in the case $d > {n/r}$.

\subsection{Stage-A} 
Stage-A of the algorithm estimates $\widehat{\Y}$ by alternating minimization of the following objective:
\begin{equation}
\widehat{\X},\widehat{\mathbf{\P}}_k =\underset{\X,\P_k \in \pi_r} {\arg \min}\,\, \lVert \Y_k - \mathbf{\P}_k \B_k \X \rVert_F ^2, \label{alt_min}
\end{equation}
$\forall k \in [n/r]$.
The minimization is initialized by setting $\widehat{\X}^{(0)}$ to the least-squares solution of the collapsed linear system of equations in \eqref{collapsed} with $\widehat{\X}^{(0)} = \widetilde{\B}^{\dagger} \widetilde{\Y}$. The update for $\widehat{\mathbf{\P}}^{(t+1)} = \text{diag}(\widehat{\P}_1,\cdots,\widehat{\P}_{n/r})$ is given by the solution to $n/r$ linear programs:
$$ \widehat{\P}_k = \underset{\P_k \in \pi_r}{\arg \min} -\tr(\widehat{\Y}_k^{(t)}\Y_k^{\intercal}\P_k),$$
$\forall k \in [n/r]$, and $\widehat{\Y}_k^{(t)} = \B_k \widehat{\X}^{(t)}$. 

The update for $\widehat{\X}$ is based on augmenting the collapsed L.S.E. 
$\left[
\begin{array}{c:c}
\widetilde{\B} & \widetilde{\Y}
\end{array}
\right]
\in \real^{{n}/{r}\times m}$ to full rank. At each iteration $t$, a measurement vector (row of matrix $\B$), $b_{p^*}^{\intercal}$, and a measurement (row of matrix $\Y$), $y^{\intercal}_{q^*}$, are appended to matrices $\widetilde{\B}^{(t)} \in \real^{n/r + t }$ and $\widetilde{\Y}^{(t)} \in \real^{n/r + t}$ respectively (line $12$ of Algorithm $1.1$). The  indices $(p^{*(t)},q^{*(t)})$ are selected (see line $11$) such that the following forward error is minimized.
\begin{equation}
(p^{*(t)},q^{*(t)}) = \underset{p \in P_k,q \in Q_k}{\arg \min}\,\, {\big \lVert \Y - \B \bigg( \big[ \frac{\widetilde{\B}^{(t)}}{b^{\intercal}_p} \big]}^{\dagger} \big[\frac{ \widetilde{\Y}^{(t)}}{y^{\intercal}_q} \big]\bigg) \big \rVert_{F} \label{forward_err},
\end{equation}
where $P_k(Q_k)$ are the sets of unmatched measurement vectors (measurements) in block $k$. The indices $p^{*(t)},q^{*(t)}$ selected at iteration $t$ are excluded from subsequent iterations so that the minimization in \eqref{forward_err} is over $n - t$ pairs of indices (line 14 of Algorithm $1.1$). Since a collapsed block by its construction depends on the rows of the block in consideration, if $r-1$ indices of a block have been matched based on the above forward error, we exclude the single remaining index in that block from subsequent iterations (lines $15$-$18$). This step is optional but it promotes the augmented system to have full row-rank. 

To augment the LSE in \eqref{collapsed} to be full-rank, we set the the number of iterations in Algorithm \ref{stage_A} equal to $d-n/r$. Algorithm \ref{stage_A} outputs $\widehat{\Y}$ and $\widehat{\mathbf{\P}}_{\mathrm{LP}}$. 
\subsection{Stage-B}
The output $\widehat{\Y}$ from Algorithm 1.1 is used to build the covariance matrix
$\widehat{\C}_{k} = \widehat{\Y}_k\widehat{\Y}_k^{\intercal}$ for the graph matching stage in Algorithm \ref{stage_B}. The output $\widehat{\mathbf{\P}}_{\mathrm{LP}}$ from Algorithm 1.1 is used to initialize the indefinite relaxation of the covariance matching program:
\begin{equation}
   \widehat{\P}_k =  \underset{\P_k \in \D_{r}} {\arg \min} -\tr (\C_{k} \P_k \widehat{\C}_{k}^{\intercal} \P_k^{\intercal}),
   \label{eq:gm_qap_1}
\end{equation}
$\forall \enskip k \in [n/r],  \widehat{\C}_{k} = \widehat{\Y}_{k}\widehat{\Y}_{k}^{\intercal}$ and ${\C}_{k}  = {\Y_{k}\Y_{k}^{\intercal}}$. Since the unknown permutation is assumed to be $r$-local, the optimization in \eqref{eq:gm_qap_1} is over the set of $r \times r$ doubly stochastic matrices. We use the Frank-Wolfe algorithm \cite{jaggi13} to minimize the objective in \eqref{eq:gm_qap_1}. A detailed description of the FW algorithm is given in\cite{fast_gm}. The doubly stochastic solution  to \eqref{eq:gm_qap_1}, $\widehat{\P}_k$, is projected onto the set of $r \times r$ permutation matrices (line $4$ of Algorithm 1.2). 

A natural question is the exactness of the indefinite relaxation of the local covariance matchings in stage-B. We will argue below that the indefinite relaxation is exact under some conditions on the covariance matrices. A similar result appears in \cite{zaslavskiy2008path} where the authors utilize graph Laplacian matrices for the indefinite graph matching problem. Let the local covariance matrices $\C_{k}$ and $\widehat{\C}_{k}$ be positive definite. Assume that there exists a permutation $\bar{\P}$ such that $\C_{k}\bar{\P} = \bar{\P}\widehat{\C}_{ k}$. Then, we claim that
\[
\underset{\P \in \D_{r}} {\arg \max} \quad \tr (\C_{k} \P \widehat{\C}_{k}^{\intercal} \P^{\intercal}) = \bar{\P}. 
\]
We prove this result as follows. Let $f(\P)$ denote the objective. We first show that $f(\P)$ is strictly convex on $\D$. A direct computation yields that, 
\[
\nabla^2 f(\P) = 2\,\widehat{\C}_{ k}\otimes \C_{k}.
\]
Using the fact that the covariances matrices are positive definite and the properties of the Kronecker product, all eigenvalues of the Hessian are positive. Therefore, $f(\P)$ is strictly convex.  Under the assumption that there exists a unique solution $\bar{\P}$ to the graph matching
problem, we have
\[
\bar{\P} = \underset{\P \in \pi_n }{\arg\min}\,\,  ||\C_{k}\P-\P\ \widehat{\C}_{k}||_{F}^{2} = \underset{\P \in \pi_n }{\arg\max}\,\,  \tr(\P^T\C_{k}^T\P\widehat{\C}_k) 
\]
A convex function over a compact convex set attains a maximizer at an extreme point\cite{bazaraa2013nonlinear}. It then follows
that 
\[
\underset{\P \in \D_n}{\arg\max}\,\,  f(\P)= \underset{\P \in \pi_n}{\arg\max}\,\,  \tr(\P^T\C_{k}\P\widehat{\C}_{k}) = \bar{\P}
\]
The implication of the aforementioned result is that the indefinite relaxation obtains the underlying solution to the graph matching problem under structural assumptions on the covariance matrices. We conjecture that if $m>r$, $\widehat{\C}_{k}, {\C}_k$ in \eqref{eq:gm_qap_1} are positive definite with high probability implying that the assumptions of the theorem hold true with the same high probability. While we have observed this empirically, a detailed analysis is left for future investigation.


\subsection{Computational complexity}

In this section, we estimate the worst-case complexity of RLUS. 

\noindent \textbf{Stage-A}: Stage-A has at most $d$ iterations. The cost of the linear program per iteration is $O(r^3)$. Minimizing the forward error is based on the pseudoinverse of an augmented matrix that varies in each iteration. For a streamlined analysis, we assume the worst-case augmentation of $\tilde{\B}$ denoted by $\bar{\B} \in \real^{d \times d}$. To compute the product $\bar{\B}^{\dagger}\Z$, for an $d\times m$ matrix $\Z$, we use the Cholesky decomposition of $\bar{\B}^{\dagger}$ followed by backward and forward substitutions. The total cost of these operations (Line 11) is $O(d^3)+O(md^2)$. The matrix product with $\B \in \real^{n\times d}$ and computing the Frobenius norm costs $O(nm)+O(ndm)$. Finally, the least squares problem to obtain $\widehat{\Y}$ (Line 24) is $O(n^3)$. With this, the worst-case complexity of stage-A is $O(d^4)+O(md^3)+O(nmd)+O(nd^2m)+O(n^3d)$. \\
\noindent \textbf{Stage-B}: Stage-B only requires forming covariance matrices and graph matching via indefinite relaxations. The cost of these operations is $O(r^3)+ O(mr^2)$. The dominating cost of $O(r^3)$ is the cost of the Hungarian algorithm which projects a given matrix onto the set of doubly stochastic matrices during the Frank Wolfe updates. \\
\noindent \textbf{Total cost}: The total worst-case complexity of the proposed algorithm is quartic with respect to $d$ and cubic with respect to $n$. In the regime of interest, $d\ll n$, the complexity in $d$ is less than cubic complexity in $n$. We remark here that the computational complexity estimate here is loose and can be sharpened using the relation of the Cholesky factorization of an augmented matrix with the original matrix \cite{gill1974methods}. 

\begin{figure*}[h!]%
\begin{subfigure}{0.24 \textwidth}
        \begin{center}
        {\includegraphics[width = 0.99 \linewidth]{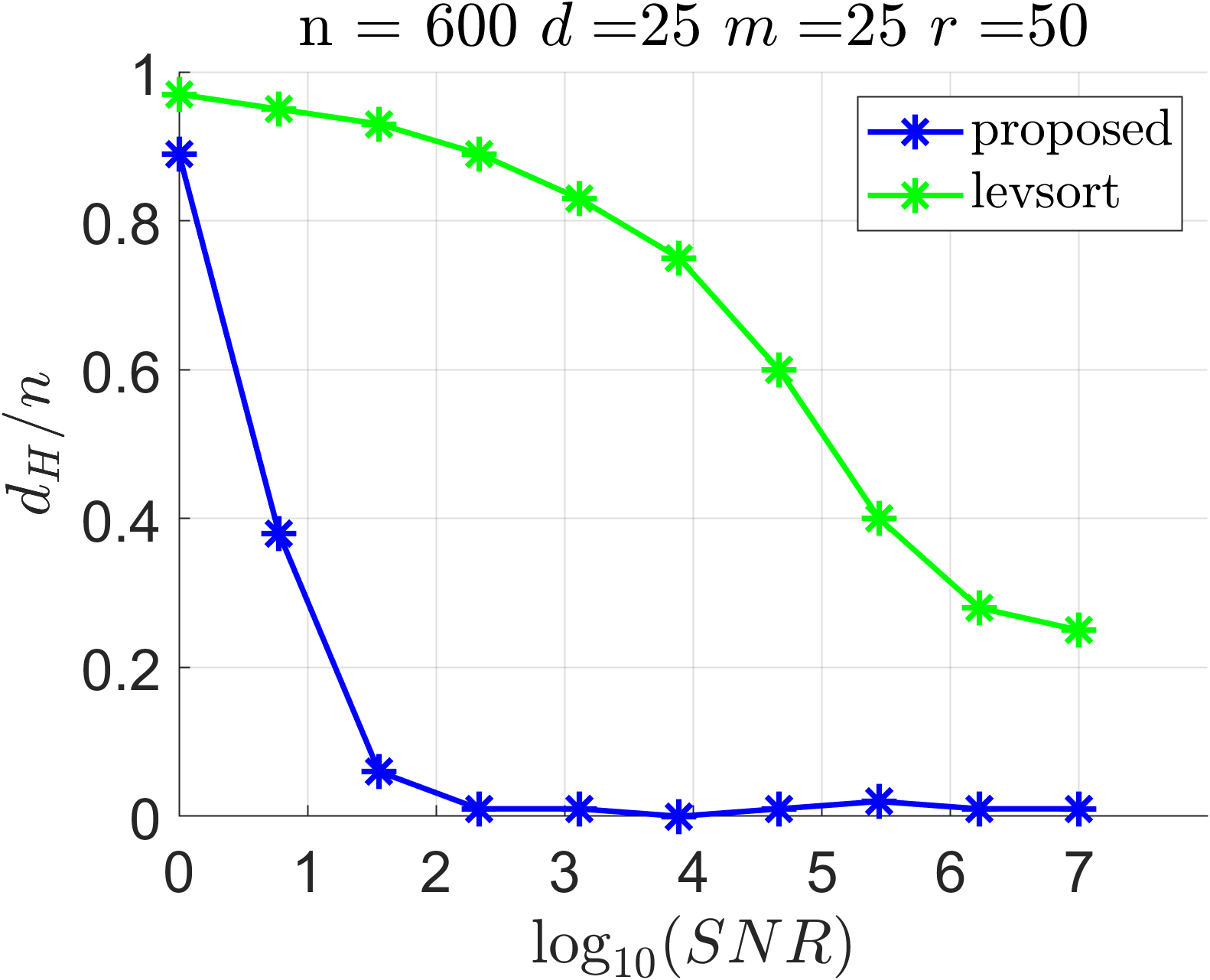}}
        \caption{}
        \label{fig:lev_comp}
        \end{center}
\end{subfigure}%
\begin{subfigure}{0.24 \textwidth}
        \begin{center}
        {\includegraphics[width = 0.99 \linewidth]{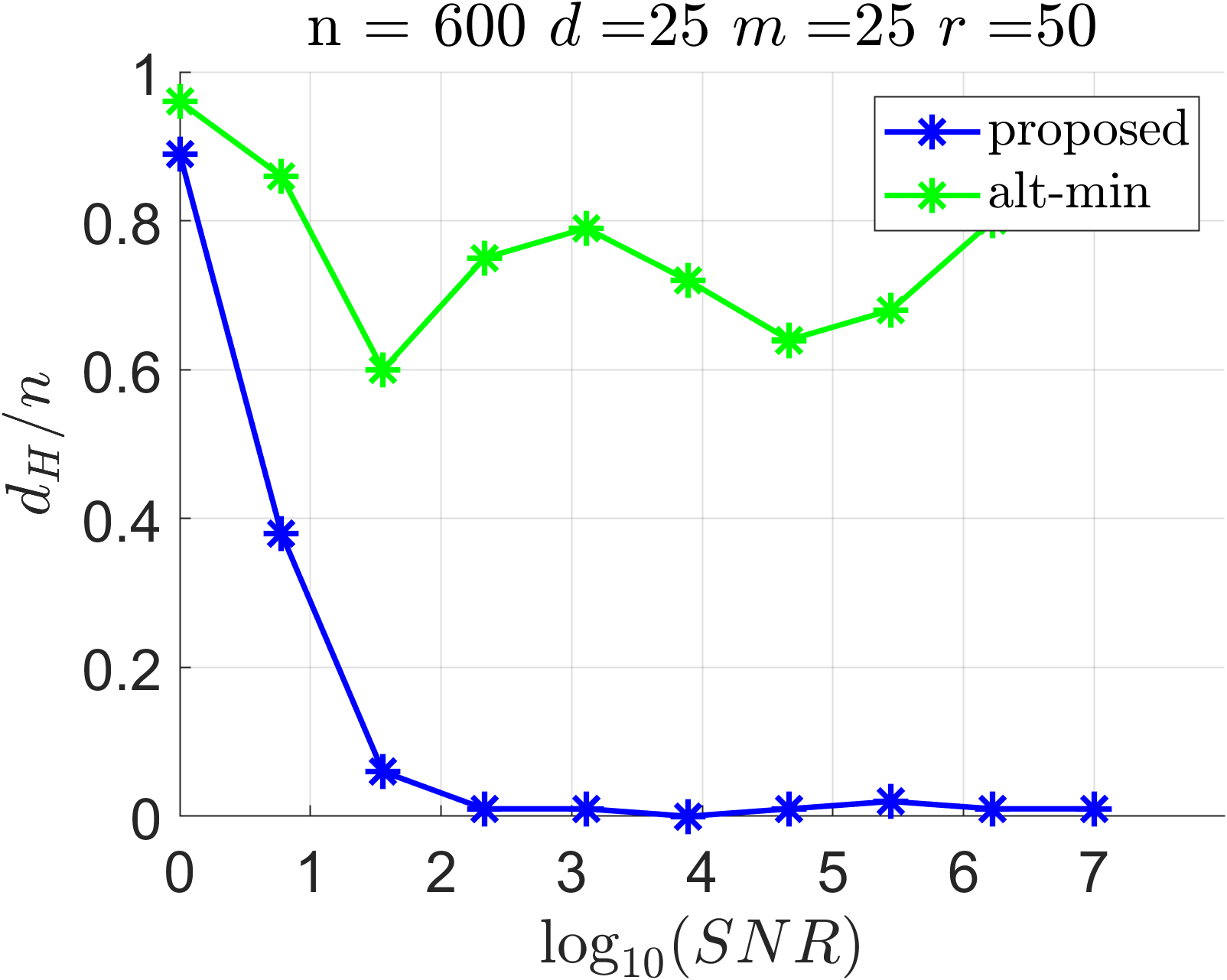}}
        \caption{}
        \label{fig:alt_min}
        \end{center}
\end{subfigure}%
\begin{subfigure}{0.24 \textwidth}
        \begin{center}
        {\includegraphics[width = 0.99 \linewidth]{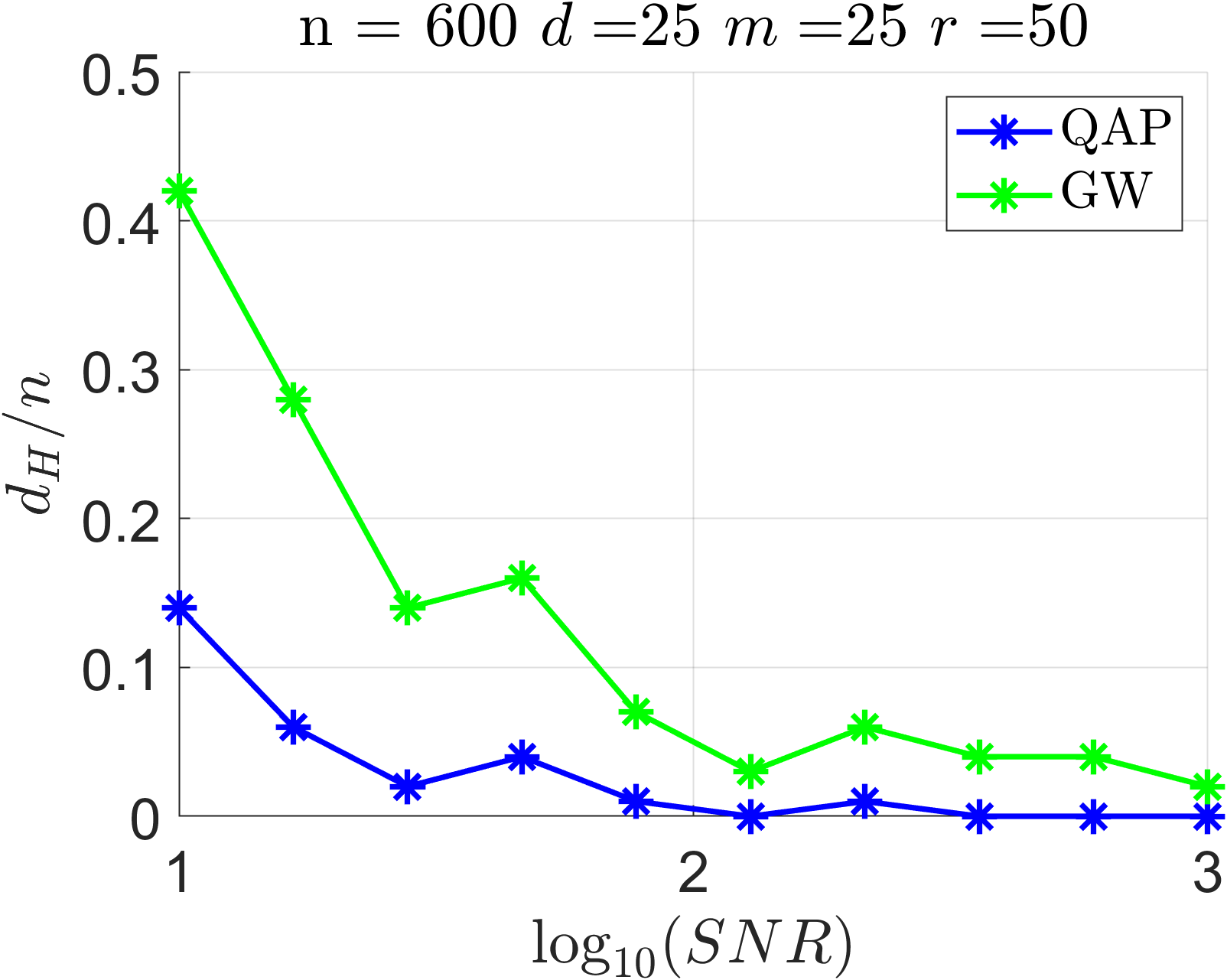}}
        \caption{}
        \label{fig:gw_vs_qap}
        \end{center}
\end{subfigure}%
\begin{subfigure}{0.24 \textwidth}
        \begin{center}
        {\includegraphics[width = 0.99 \linewidth]{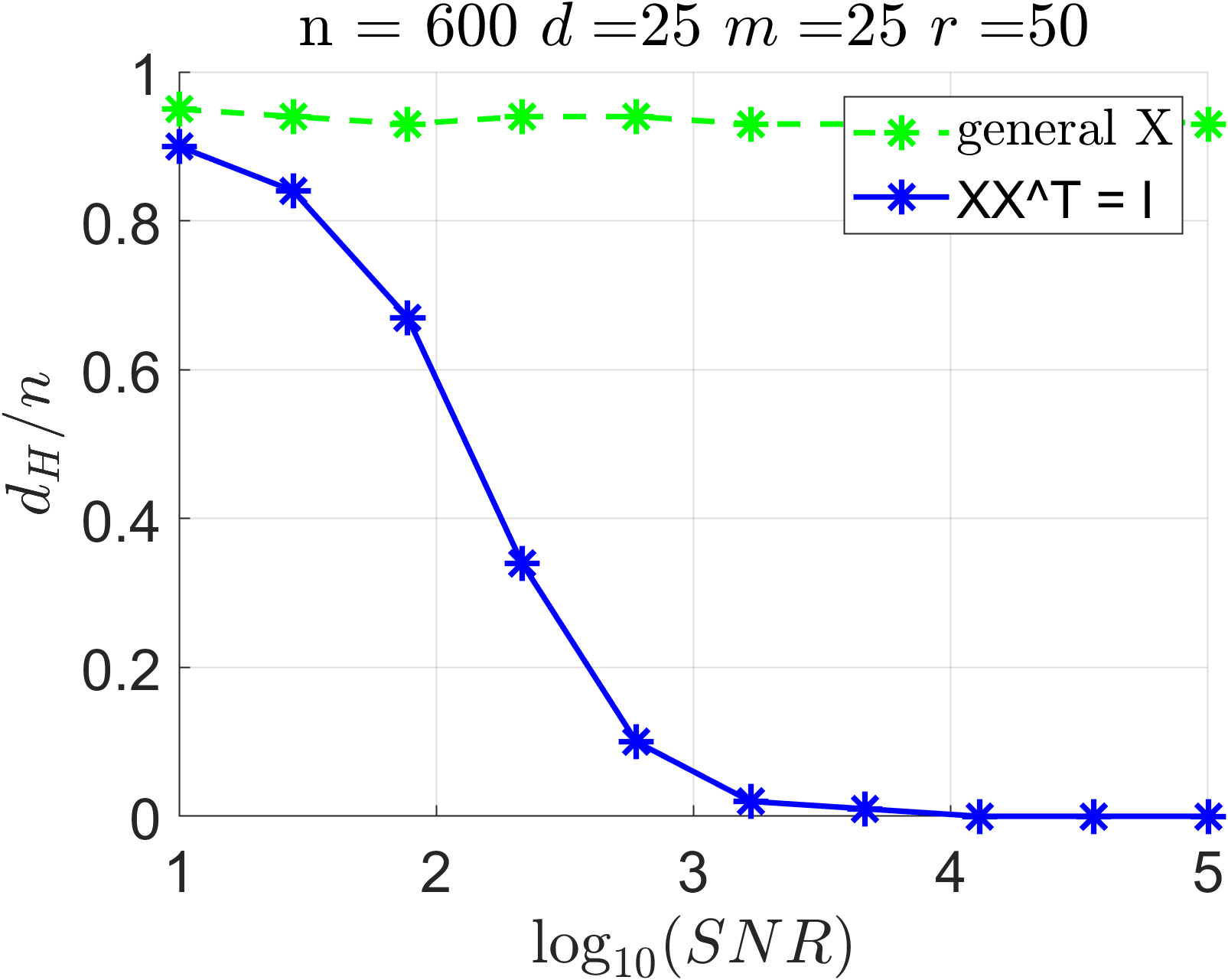}}
        \caption{}
        \label{fig:orth_X}
        \end{center}
\end{subfigure}
\hfill%
\caption{$\Y = \boldsymbol{\P}^{*} \B_{n \times d} \X^*_{d \times m} + \mathbf{N}$. Figure plots the fractional Hamming distortion, $d_H(\mathbf{\P}^*,\widehat{\mathbf{\P}})/n$,  against $\textrm{SNR}$. }
\label{fig:benchmark}
\end{figure*}
\begin{figure*}[h!]%
\begin{subfigure}{0.24 \textwidth}
        \begin{center}
        {\includegraphics[width=\linewidth]{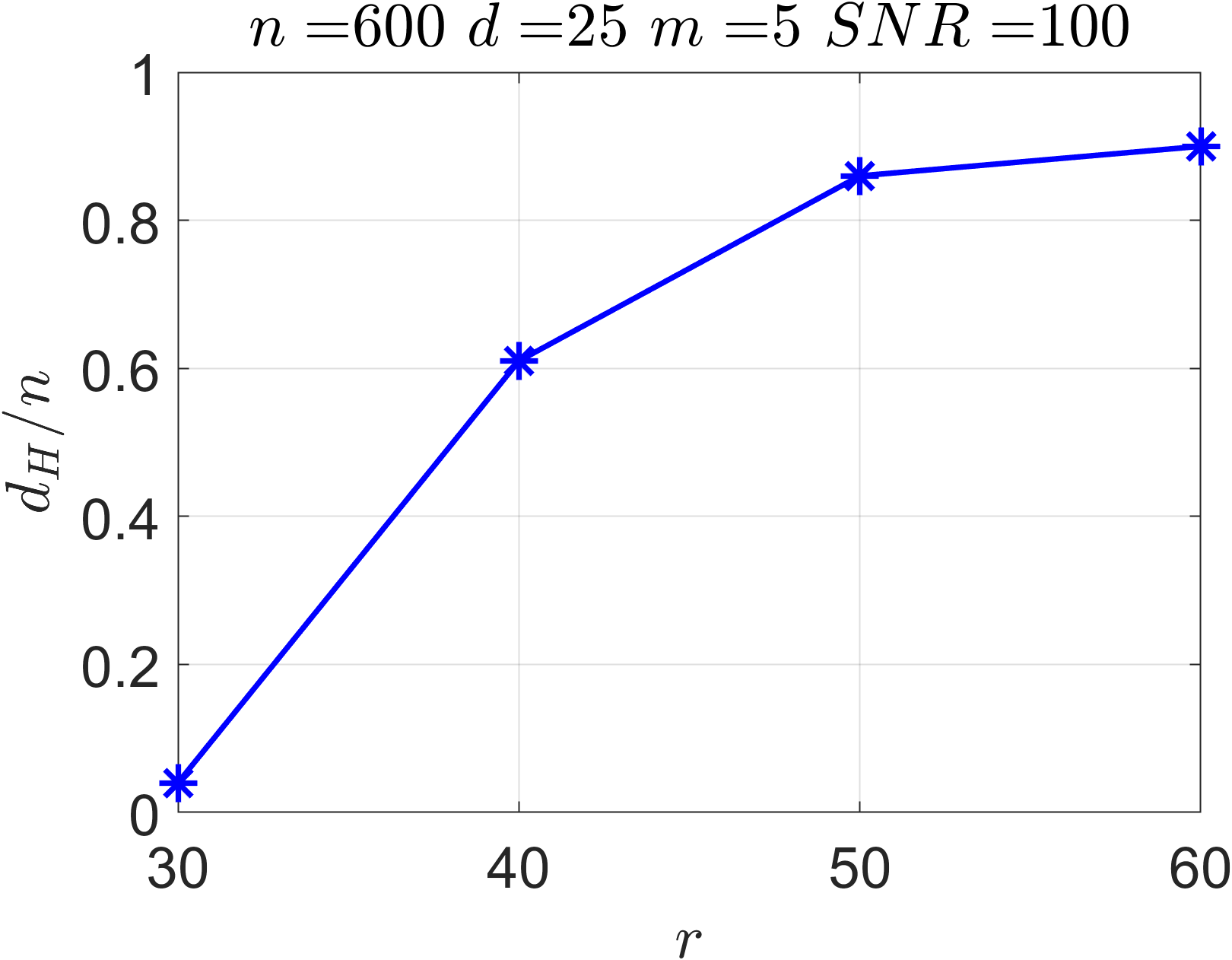}}
        \caption{}
        \label{fig:m_small}
        \end{center}
\end{subfigure}%
\begin{subfigure}{0.24 \textwidth}
        \begin{center}
        {\includegraphics[width = \linewidth]{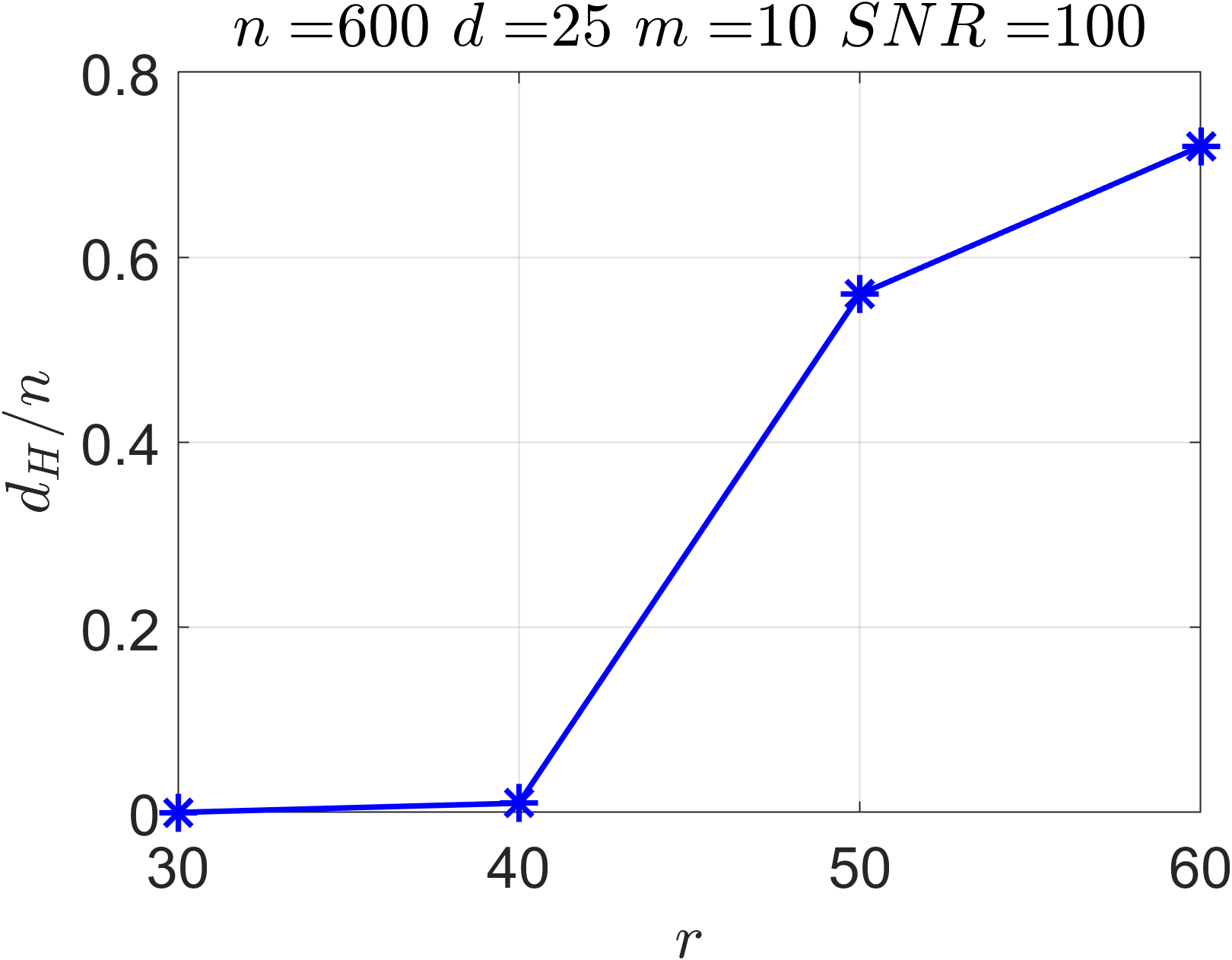}}
        \caption{}
        \end{center}
\end{subfigure}%
\begin{subfigure}{0.24 \textwidth}
        \begin{center}
        {\includegraphics[width = \linewidth]{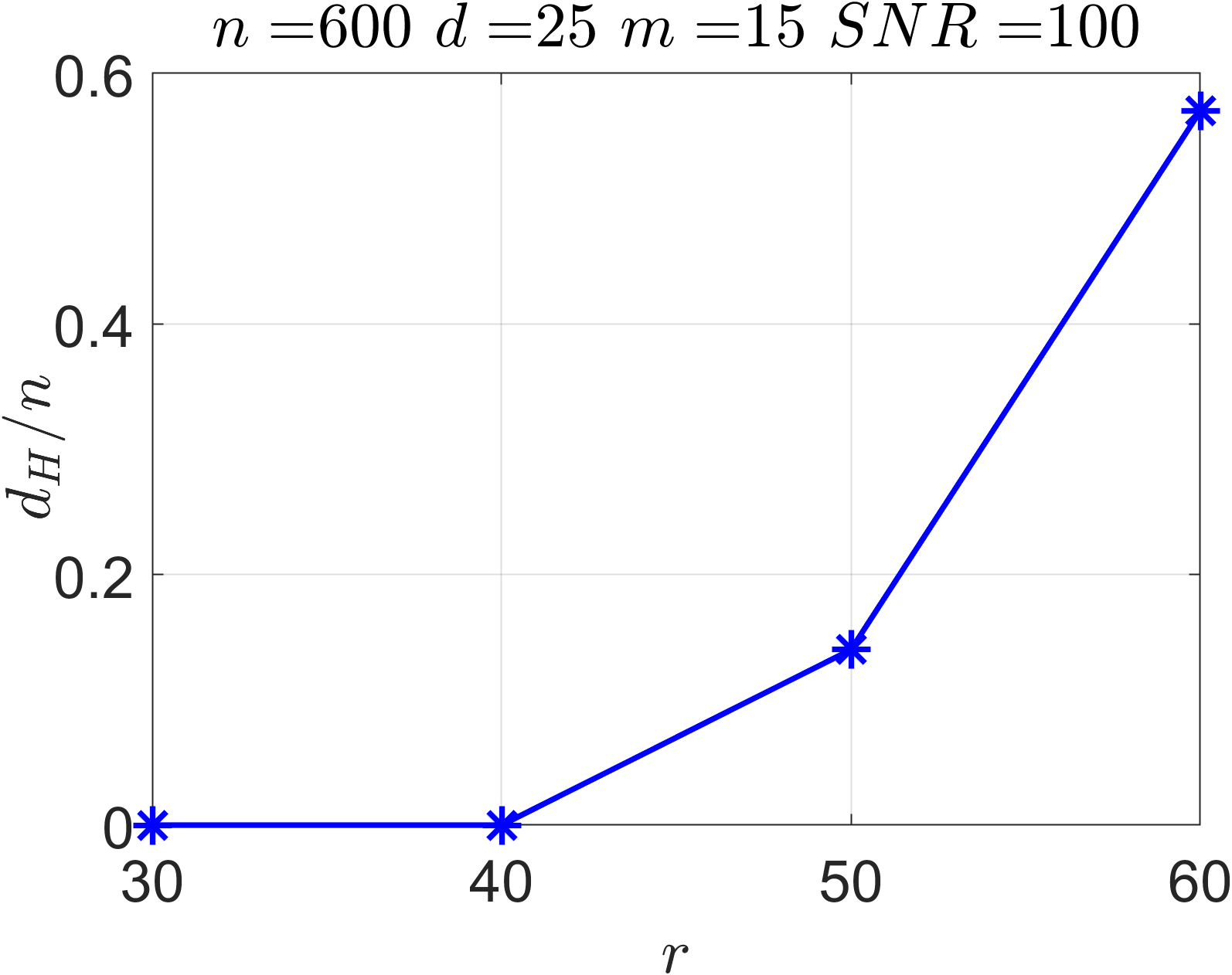}}
        \caption{}
        \end{center}
\end{subfigure}%
\begin{subfigure}{0.24 \textwidth}
        \begin{center}
        {\includegraphics[width = \linewidth]{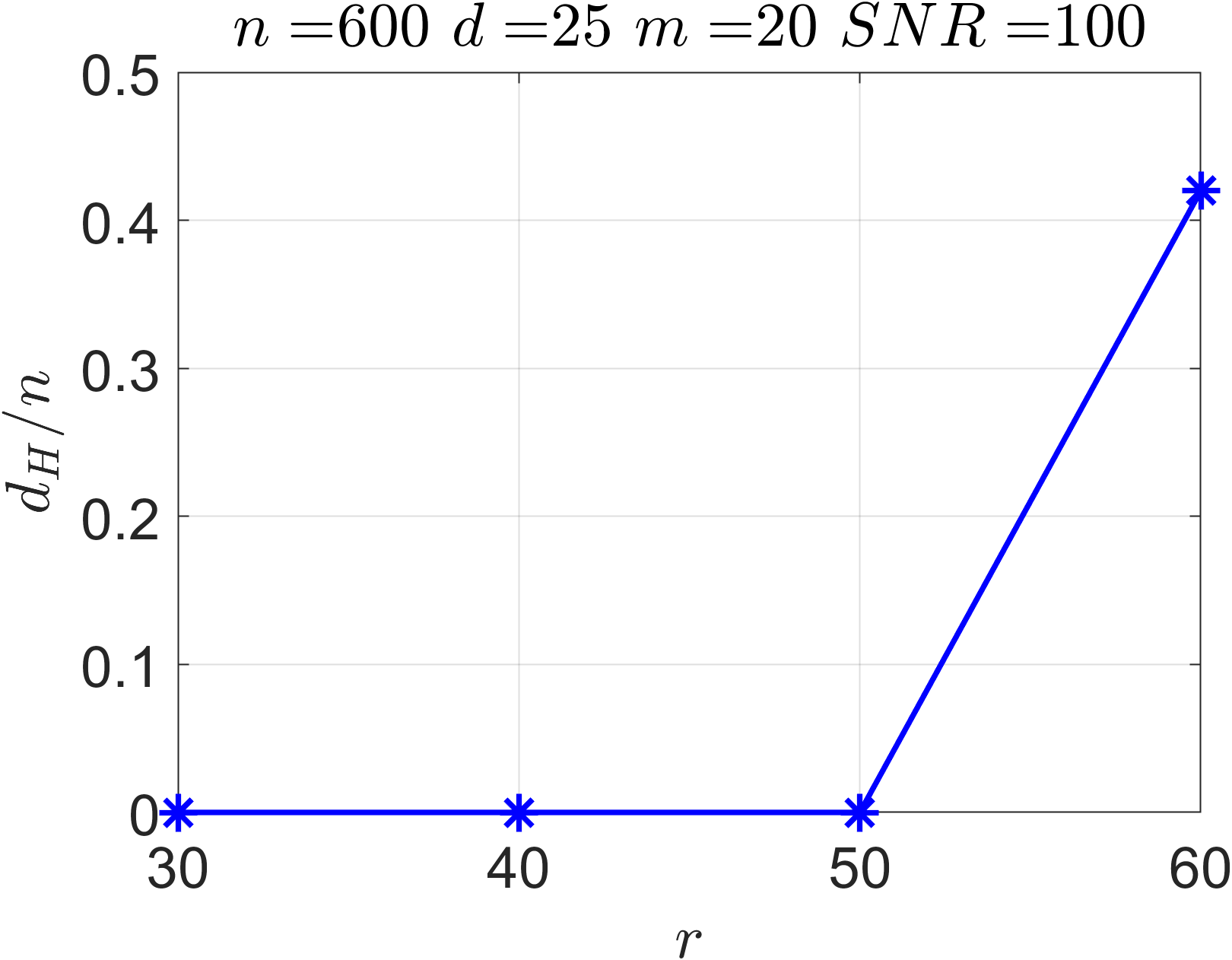}}
        \caption{}                
        \label{fig:m_big}
        \end{center}
\end{subfigure}%
\hfill%
\begin{subfigure}{0.24 \textwidth}
        \begin{center}
        {\includegraphics[width = \linewidth]{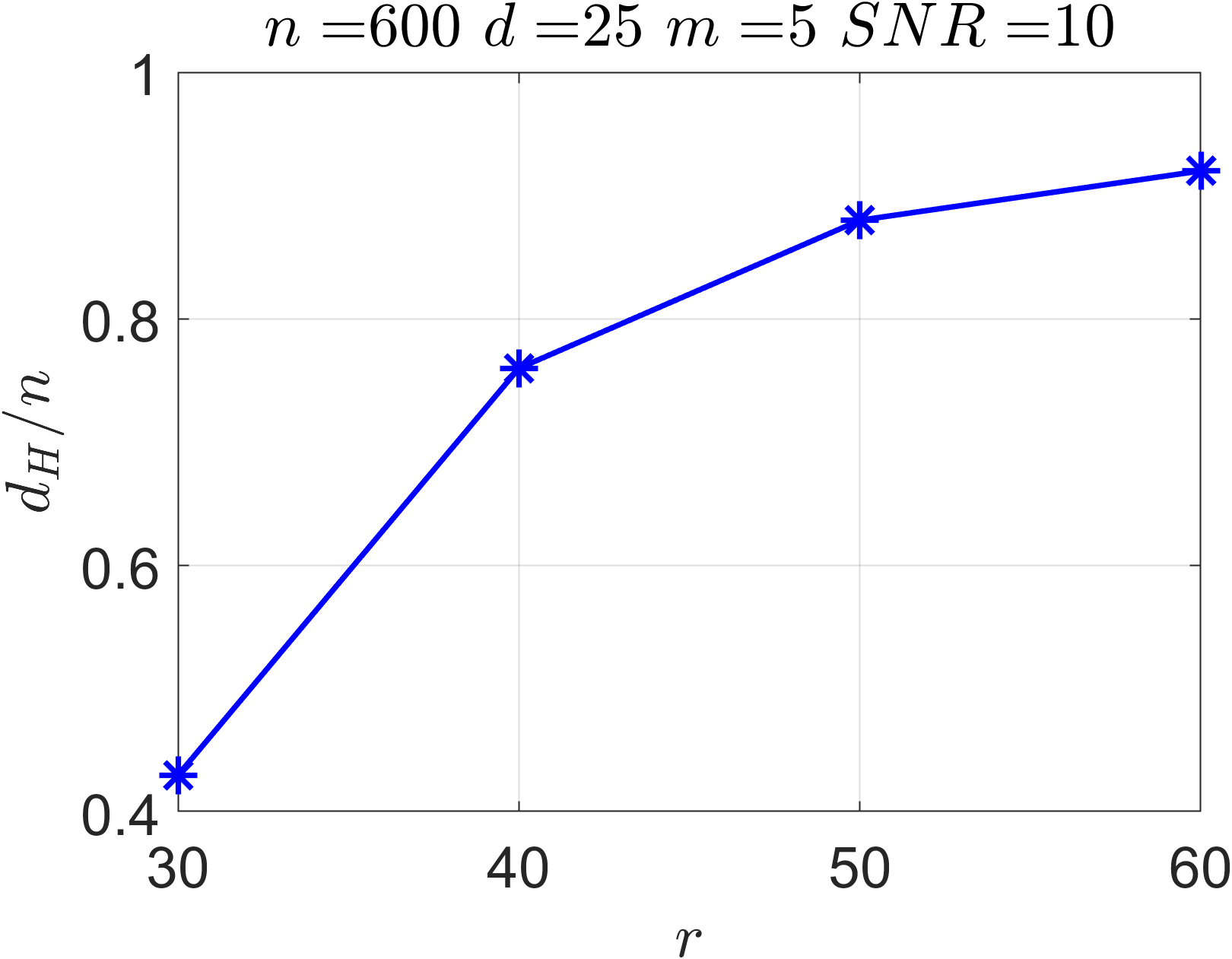}}
        \caption{}
        \end{center}
\end{subfigure}%
\begin{subfigure}{0.24 \textwidth}
        \begin{center}
        {\includegraphics[width = \linewidth]{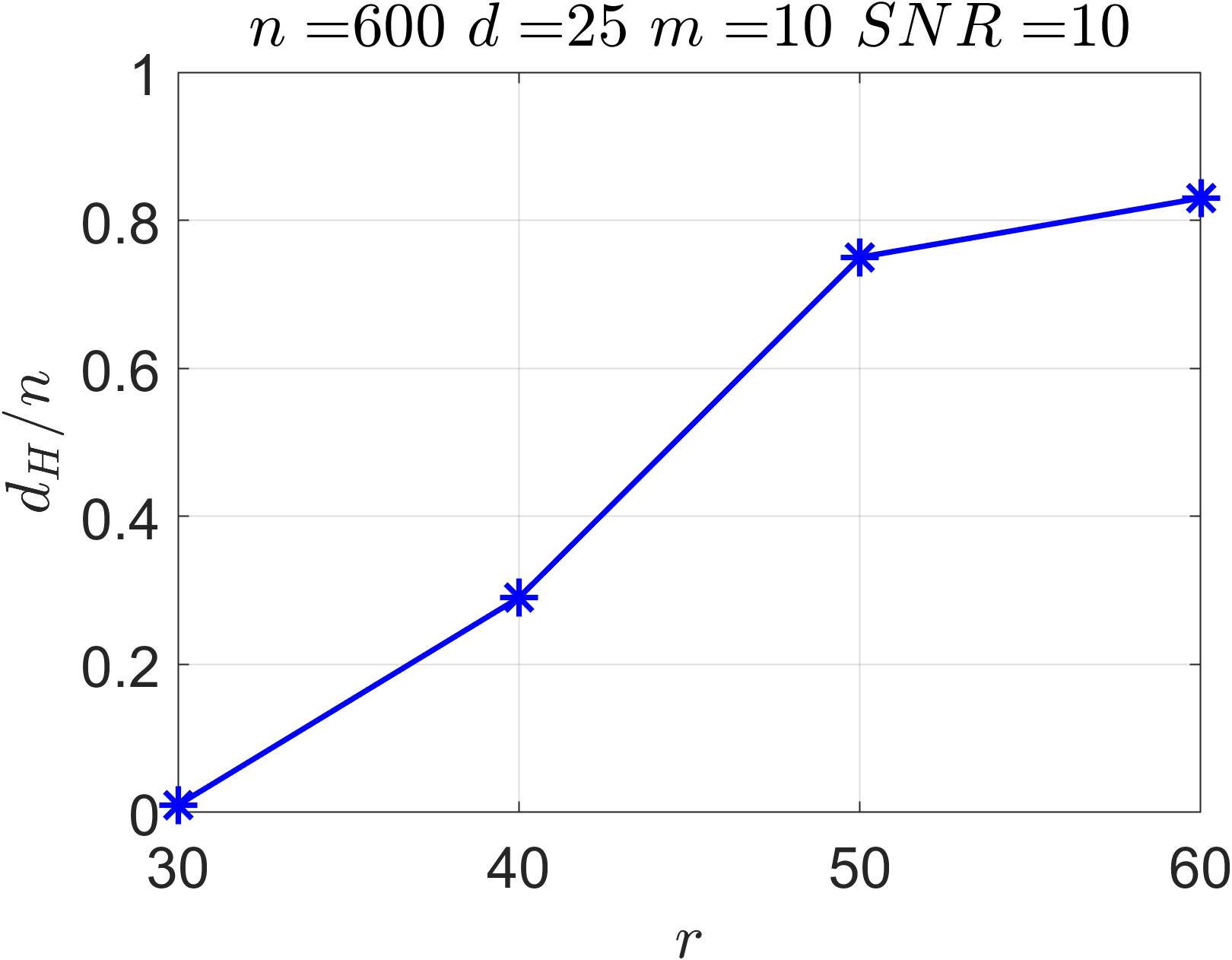}}
        \caption{}
        \end{center}
\end{subfigure}%
\begin{subfigure}{0.24 \textwidth}
        \begin{center}
        {\includegraphics[width= \linewidth]{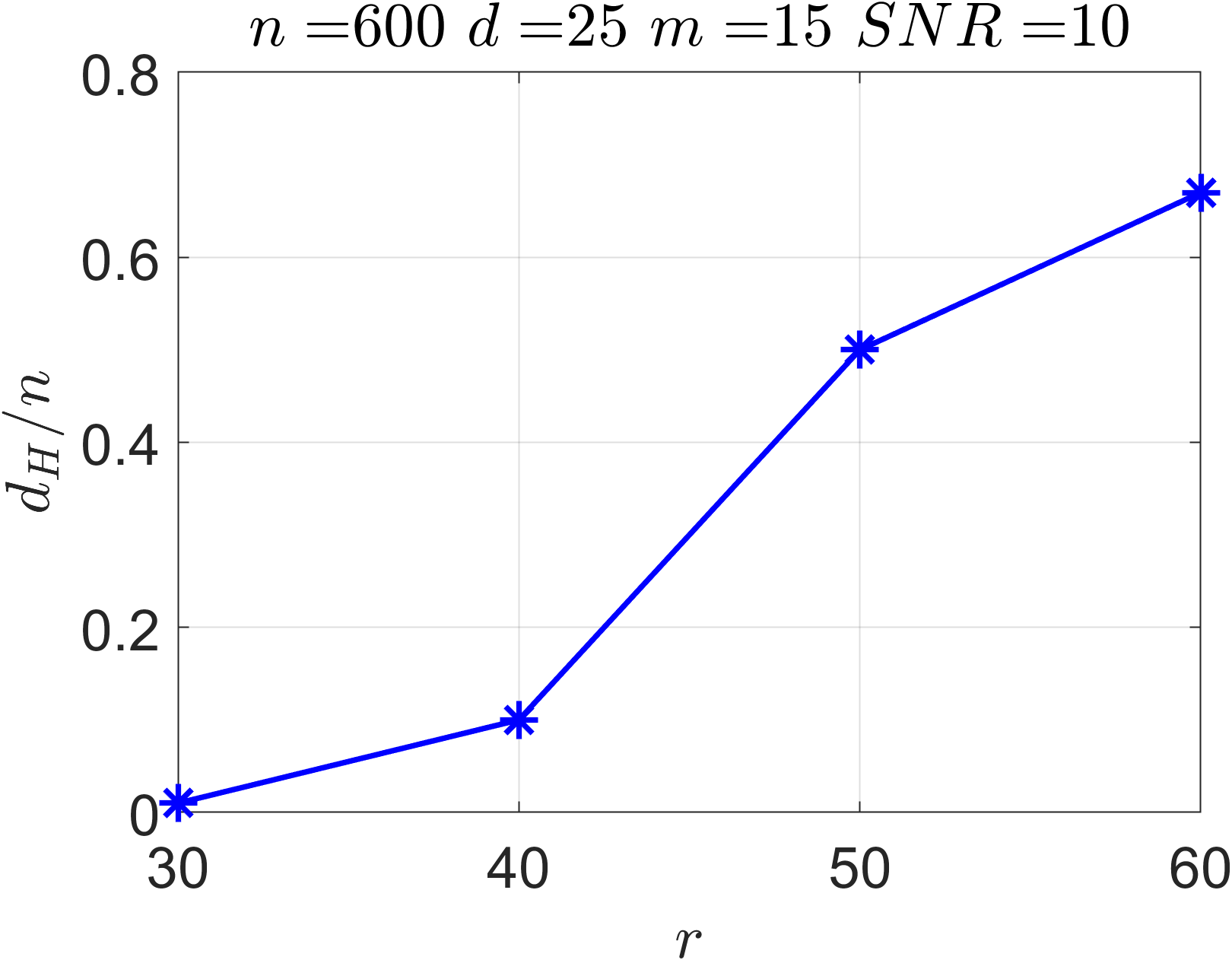}}
        \caption{}
        \end{center}
\end{subfigure}%
\begin{subfigure}{0.24 \textwidth}
        \begin{center}
        {\includegraphics[width =  \linewidth]{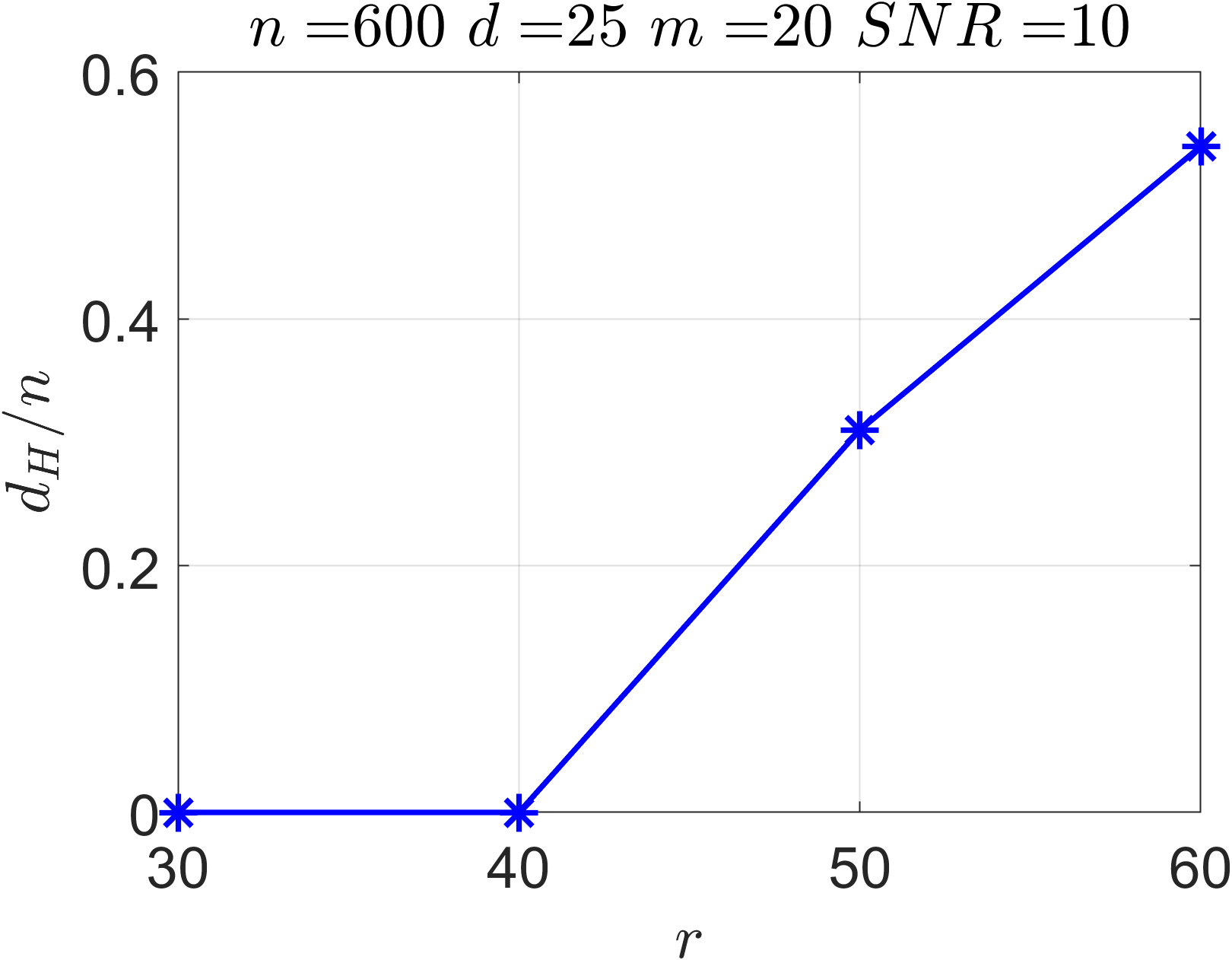}}
        \caption{}
        \end{center}
\end{subfigure}
\hfill%
\caption{$\Y = \boldsymbol{\P}^{*} \B_{n \times d} \X^*_{d \times m} + \mathbf{N}, d=25$. Figure plots the fractional Hamming distortion, $d_H(\mathbf{\P}^*,\widehat{\mathbf{\P}})/n$,  against block size $r$ for $\textrm{SNR} \in \{10,100\}$. }
\label{fig:d_25}
\end{figure*}
\begin{figure*}[h!]%
\begin{subfigure}{0.24 \textwidth}
        \begin{center}
        {\includegraphics[width =  \linewidth]{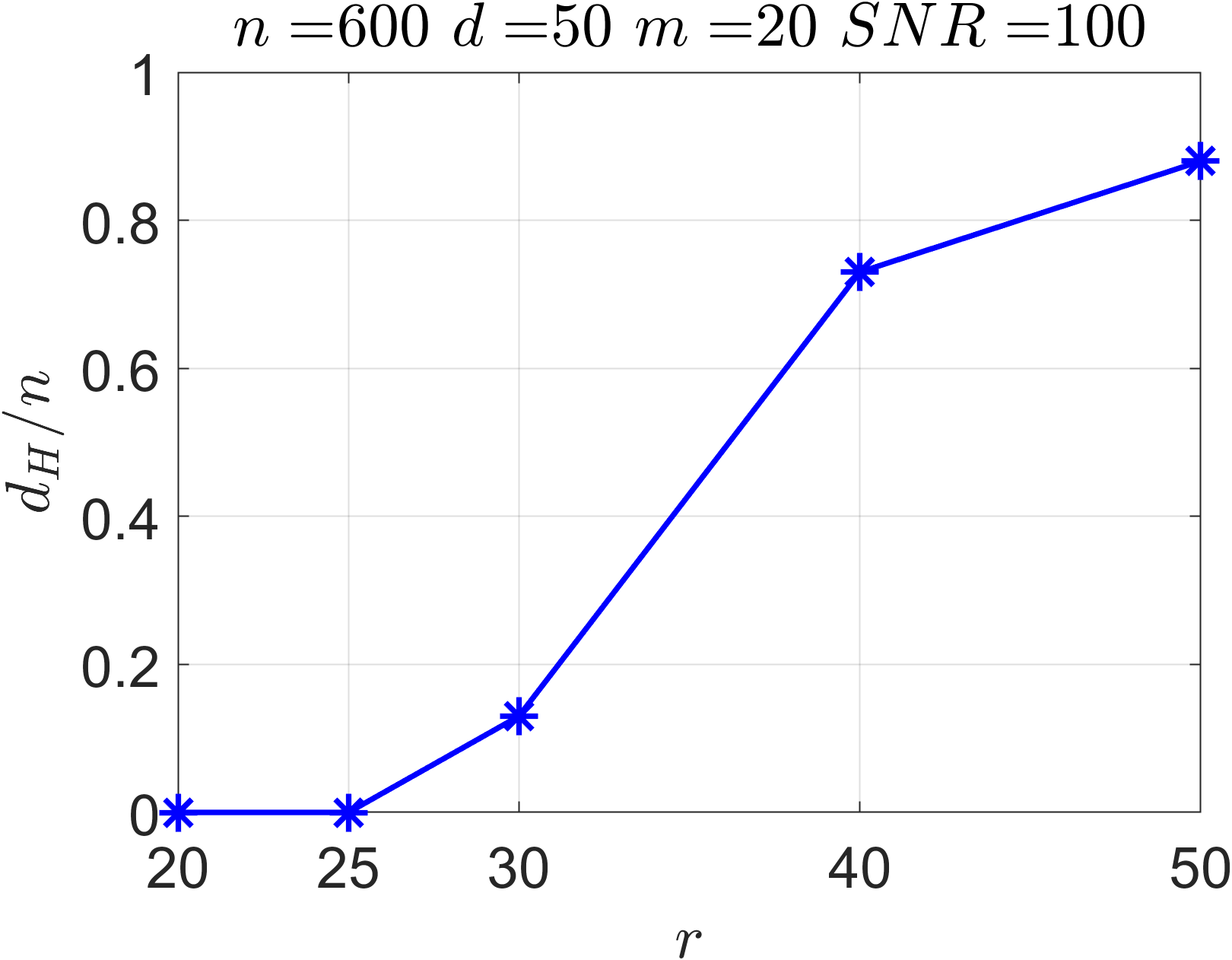}}
        \caption{}
        \end{center}
\end{subfigure}%
\begin{subfigure}{0.24 \textwidth}
        \begin{center}
        {\includegraphics[width =  \linewidth]{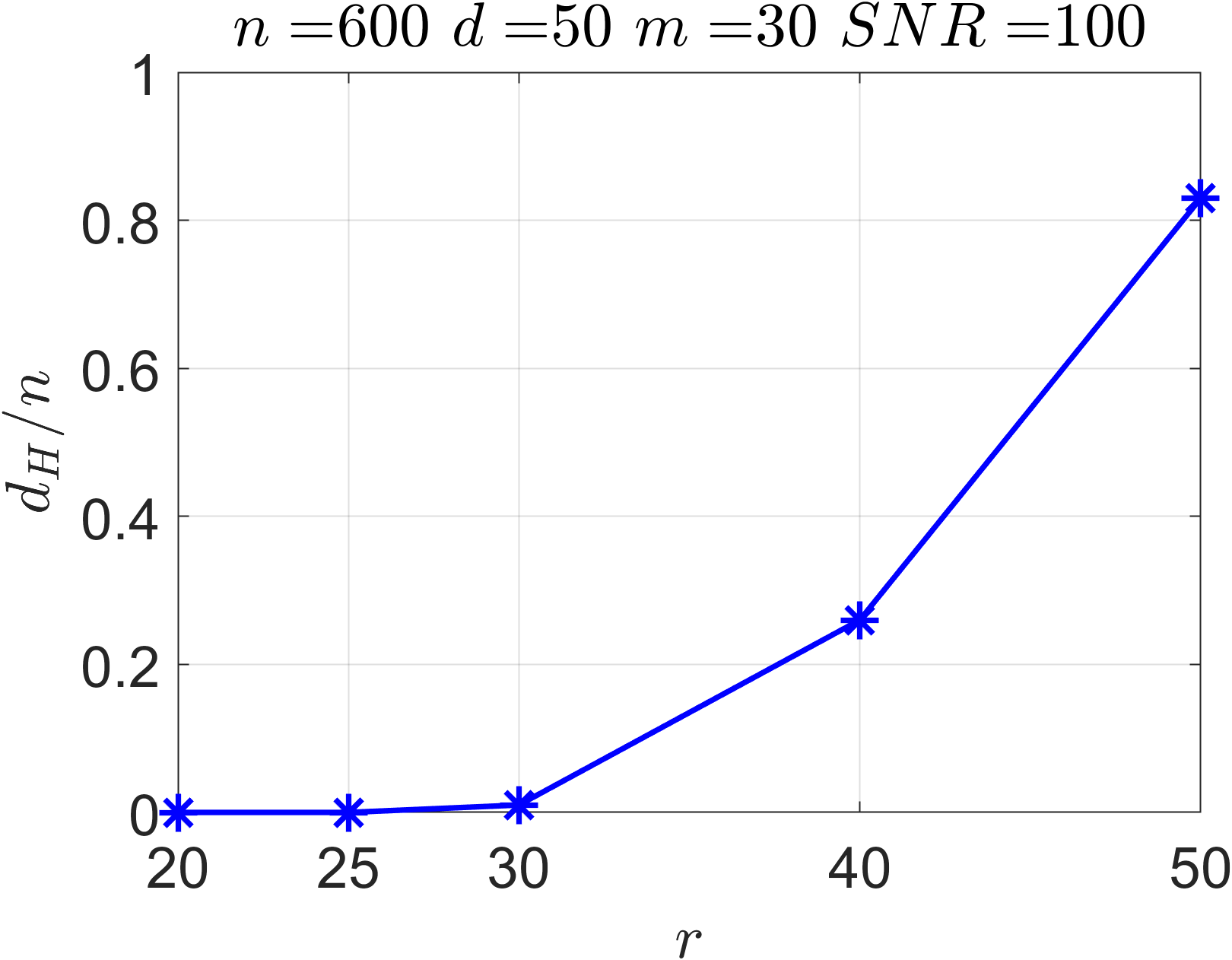}}
        \caption{}
        \end{center}
\end{subfigure}%
\begin{subfigure}{0.24 \textwidth}
        \begin{center}
        {\includegraphics[width =  \linewidth]{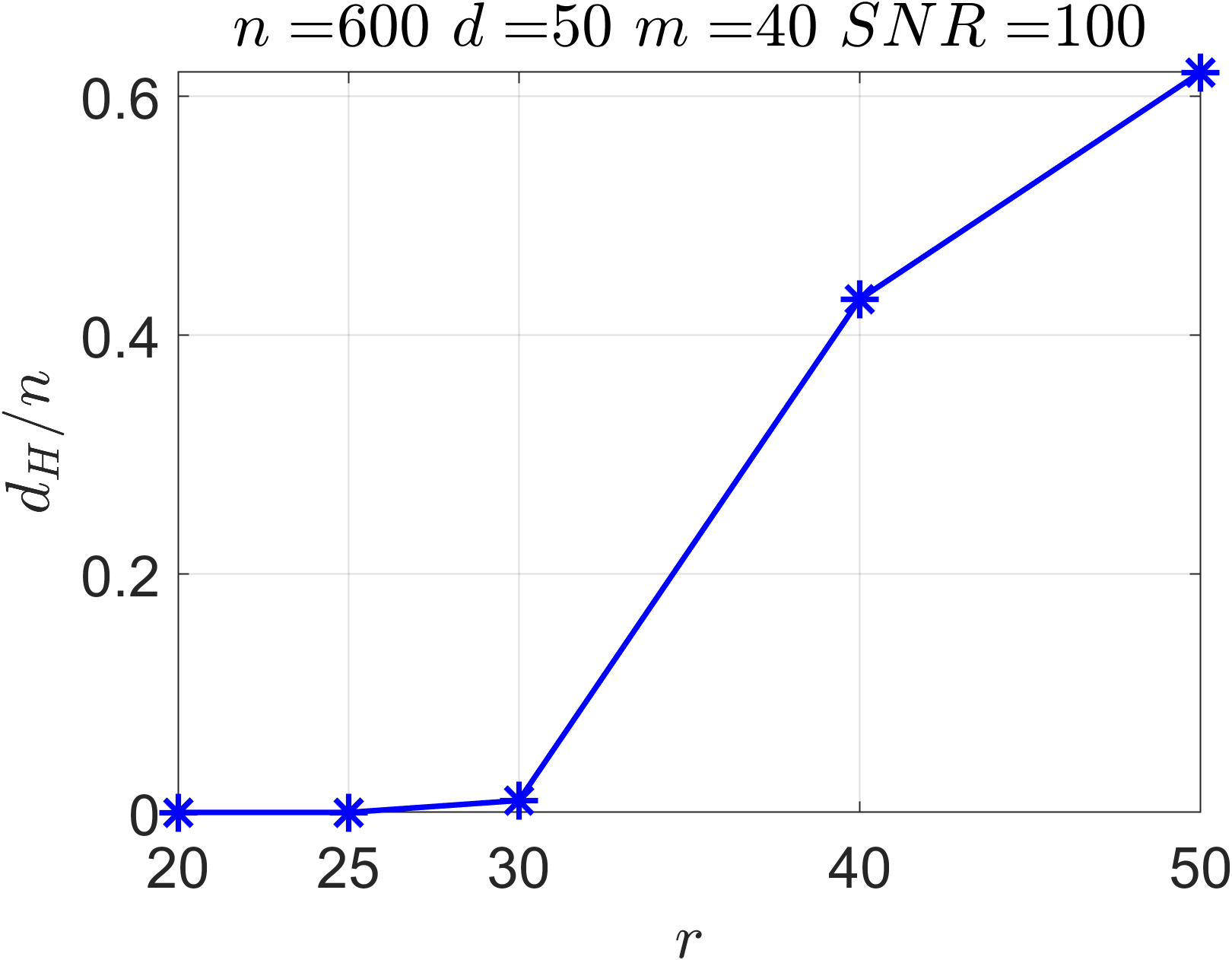}}
        \caption{}
        \end{center}
\end{subfigure}%
\begin{subfigure}{0.24 \textwidth}
        \begin{center}
        {\includegraphics[width =  \linewidth]{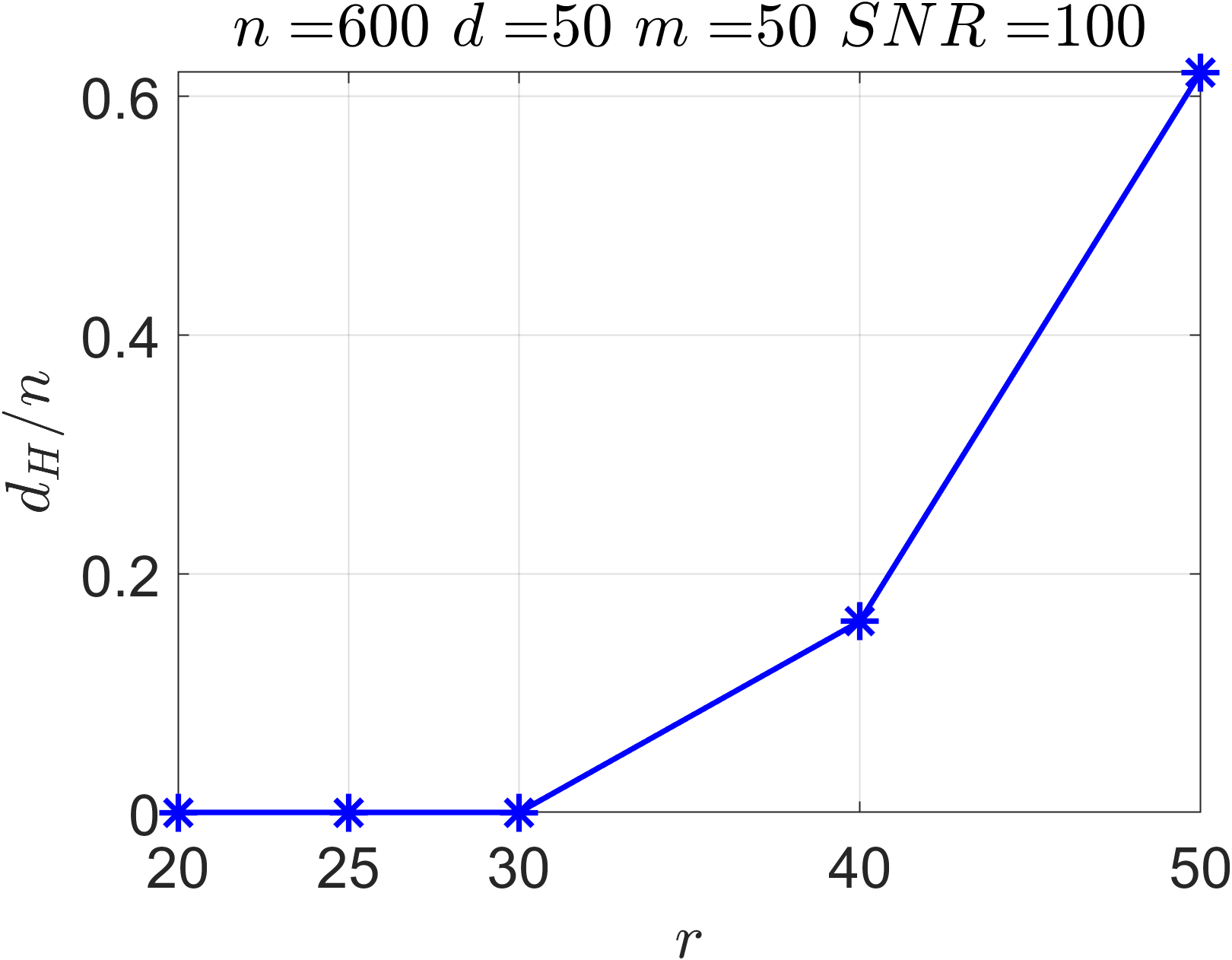}}
        \caption{}                
        \end{center}
\end{subfigure}
\hfill%
\caption{$\Y = \boldsymbol{\P}^{*} \B_{n \times d} \X^*_{d \times m} + \mathbf{N}, d=50$. Figure plots the fractional Hamming distortion, $d_H(\mathbf{\P}^*,\widehat{\mathbf{\P}})/n$,  against block size $r$ for $\textrm{SNR} = 100$. }
\label{fig:d_50}
\end{figure*}
\section{Results}
\label{sec:results}
In this section, we show the performance of RLUS, implemented in MATLAB, on synthetic data sets. All of our numerical experiments are evaluated with respect to the ground truth permutation using the fractional Hamming distortion. 

\noindent \textbf{Synthetic data}: 
For the model in \eqref{eq:local}, we draw entries of the measurement matrix $\mathbf{B} \in \mathbb{R}^{n \times d}$ and the signal $\mathbf{X}^*\in \mathbb{R}^{d \times m}$ i.i.d from the standard normal distribution. The additive noise $\mathbf{N}$ is drawn i.i.d from Gaussian distribution $\mathcal{N}(0,\sigma^2)$, and the permutation ${\P}^*$ is picked uniformly at random from the set $\pi_{n,r}$. Following other works in the unlabeled sensing literature \cite{slawski_two_stage,rahmani,zhang2019permutation}, the signal to noise ratio is defined as  \begin{align} \textrm{SNR} =\frac{\lVert \mathbf{X}\rVert_F^2}{m\sigma^2}. \end{align}
All results are averaged over $25$ Monte Carlo runs.

\noindent \textbf{Baselines}: We compare the proposed approach to $4$ baseline algorithms:

\textit{i) Subspace matching}: The \textit{levsort} algorithm in \cite{LEVSORT} attempts to recover the unknown permutation by matching the subspace spanned by $\mathbf{U}_B$, the left singular vectors of the measurement matrix $\B$, to $\mathbf{U}_Y$. For $m = d$ views in the noiseless setting, the \textit{levsort} algorithm recovers the unknown permutation exactly.  

\textit{ ii) Alt-min}:  We compare the proposed approach against an alternating minimization approach for the biconvex objective in \eqref{eq:LP}. The updates on $\widehat{\P},\widehat{\X}$ are given as:
\begin{align*}
    \widehat{\X}^{(t)} &= \B^{\dagger} (\widehat{\P}^{(t){\intercal}} \Y) \tag{alt-min}\\
    \widehat{\P}_k^{(t+1)} &= \underset{\P_k \in \pi_r}{\arg \min} -\tr (\B\widehat{\X}^{(t)}\Y \P_k)
\end{align*}
$\forall  k \in [n/r]$, and $\widehat{\P}^{(0)}_k = \mathbf{I}_{r \times r}$, $\widehat{\P} = \textrm{diag}(\widehat{\P}_1,\cdots,
\widehat{\P}_{n/r})$.

\textit{iii)  Covariance matching via GW alignment}: We compare the QAP based graph matching in Algorithm \ref{algo:gm} to the Gromov-Wasserstein alignment. 
Given two graphs with adjacency matrices $\A$ and $\B$, the Gromov Wasserstein alignment consists in minimizing the square of the Gromov-Wasserstein distance \cite{memoli2011gromov}, $\textrm{GW}_2^2(\A,\B)$, between them: 
\begin{equation} \textrm{GW}^2(\A,\B) = \min_{\P \in  \mathbb{D}_n}\sum_{p',q'} \sum_{p,q} {\big(\B_{p'q'} - \A_{pq} \big)}^2 \P_{p'p} \P_{q'q}
\label{GW_distance}
\end{equation} 
 The GW alignment problem in \eqref{GW_distance} is a non-convex quadratic optimization problem. We use the algorithm proposed in \cite{peyre2016gromov} to solve \eqref{GW_distance} with entropic regularization:
  $$\textrm{GW}^2_{\epsilon}(\A,\B) =  \textrm{GW}^2(\A,\B) - \epsilon H(\P),$$
  where entropy $H(\P) = - \sum P_{pq}(\log(P_{pq})-1)$, and $\epsilon$ is the amount of entropic regularization. A higher value of $\epsilon$ (high entropic regularization) returns a diffused/soft assignment between the weighted adjacency matrices $\A,\B$. In contrast, lower values of $\epsilon$ return more precise assignments.

\textit{iv) Covariance matching for orthonormal $\X^*$}: For an orthonormal signal matrix, $\X^*(\X^{*})^{\intercal} = \mathbf{I}$, as considered in \cite{spherical}, we recover $\widehat{\P}^*$ by minimizing the objective in \eqref{orth_X}.

\subsection{Comparisons with baselines}
For all comparisons in  {Fig. \ref{fig:benchmark}}, we set the block size $r= 50$, number of measurements $n=600$, views $m=d=25$, and plot the fractional Hamming distortion $d_H/n$ against $\textrm{SNR}$. 

\noindent{\textbf{Comparison with levsort}}: In Fig. \ref{fig:lev_comp}, we compare our algorithm to the subspace matching algorithm \textit{levsort} in \cite{LEVSORT}. We observe that the algorithm in \cite{LEVSORT} is not robust to noise as the distortion in the estimate returned by \textit{levsort} (green) is high  even for diminishing levels of noise power. This is in contrast to the proposed algorithm which returns $d_H(\widehat{\P},\P^*) = 0$ for $\textrm{SNR} = 1000$. 

\noindent{\textbf{Comparison with alt-min}}: In Fig. \ref{fig:alt_min}, we observe that the alternating minimization approach in \eqref{alt_min} fails to recover the unknown permutation $\widehat{\P}^*$. This is because, unlike the models considered in \cite{slawski_two_stage,rahmani,zhang2019permutation}, the $r$-local model in this work does not assume partially shuffled measurements. In particular, the initialization $\widehat \P^{(0)} = \mathbf{I}_{r \times r}$ is suitable only when the measurements are partially shuffled.

\noindent{\textbf{Comparison with GW alignment}}: In Fig. \ref{fig:gw_vs_qap}, we observe that the proposed QAP alignment outperforms GW, especially in the low $\textrm{SNR}$ regime. For both QAP and GW, we let $\A = \Y_k  \Y_k^{\intercal}$ and $\B = \widehat\Y_k \widehat \Y_k^{\intercal}$.  We set the amount of entropic regularization for GW to the lowest value that avoided numerical instability in the experiments. 

\noindent{\textbf{$\X\X^{\intercal} = \mathbf{I}$}}: In Fig. \ref{fig:orth_X}, we observe that, for orthonormal $\X^*$  (blue), the unknown permutation $\P^*$ can be recovered by aligning the covariance blocks $\B_k\B_k^{\intercal}$ to $\Y_k\Y_k^{\intercal}$. However, this approach fails for general $\X$ (green). The results corroborate the claim in section \ref{subsec:orth_X} that, for orthonormal $\X^*$ considered in \cite{spherical}, the permutation can be recovered without estimating the unknown signal $\X^*$.

\subsection{Other simulations}
In Figures \ref{fig:d_25}, \ref{fig:d_50}, we plot the fractional Hamming distortion $d_H{(\widehat{\P},\P^*)}/n$  against block size $r$. 

For the simulations in Fig. \ref{fig:d_25}, we set signal dimension $d=25$. For $m \in \{5,10,15,25\}$, we plot $d_H{(\widehat{\P},\P^*)}/n$  against $r \in \{30,40,50,60\}$. We observe the following:

\noindent \textbf{i)}  Hamming distortion in the returned estimate is lower for small values of $r$ and increases with increasing $r$. This is because for fixed $n$, as $r$ increases, the collapsed system in \eqref{collapsed} becomes increasingly underdetermined. Secondly, the size of each local permutation, $r!$, increases with increasing $r$.

\noindent \textbf{ii)} Permutation recovery is significantly improved with increasing views. For example, for $r=50$, $m=5$ in Fig. \ref{fig:m_small}, the Hamming distortion is nearly $1$. In contrast, for $m=20$ views in Fig. \ref{fig:m_big}, the Hamming distortion is $0$.  As shown in \cite{zhang2019permutation}, the necessary $\textrm{SNR}$ required for recovering $\P^*$ reduces from $\textrm{SNR} = \Omega (n^{2})$ for the single-view case to $\textrm{SNR} \simeq n^{1/\rho(\X)}$, for the multi-view setting \footnote{Eq. (7) in \cite{zhang2019permutation}}.  The stable rank $\rho(\X)$ of matrix $\X$ is defined as the ratio $\sum \sigma^2(\X)/\sigma^2_{\textrm{max}}(\X)$, where $\sigma(\X)$ are the singular values of $\X$. For $\X \in \mathbb{R}^{d \times m}$, the stable rank is upper bounded by  $\rho(\X) \leq \textrm{rank}(\X) =  \min(m,d)$. Increasing the number of views $m$ increases the stable rank of $\X$ and reduces the necessary $\textrm{SNR}$ for permutation recovery. In Fig.\ref{fig:d_25} (e)-(h), we decrease $\textrm{SNR}$ from $\textrm{SNR}=100$ in Fig. \ref{fig:d_25} (a)-(d) to $\textrm{SNR} = 10$.

For the simulations in Fig. \ref{fig:d_50}, we set dimension $d=50$, $\textrm{SNR}=100$, and plot the Hamming distortion against $r\in\{20,25,30,40,50\}$. We can draw similar conclusions as from Fig. \ref{fig:d_25}. The distortion in the estimated permutation increases with increasing $r$ and decreases with increasing $m$. However, for the same block size $r$, the distortion in $\widehat \P$ is higher for $d=50$ than for $d=25$. The collapsed system in \eqref{collapsed} used to initialize our algorithm is underdetermined by $d-n/r$ measurements. A higher value of $d$ corresponds to a poor initialization of the non-convex optimization problem in Stage-A. Since the permutation estimate $\widehat \P$ in stage-B is obtained from aligning $\widehat{\C}_k$ to $\C_k$, the distortion in this estimate is higher when $\widehat{\C}_{k}$ is not estimated well. 
\section{Conclusion}
\label{sec:conclude}
We study a linear inverse problem with scrambled measurements well known as the unlabeled sensing problem. In this paper, we focus on a novel setup of this problem, the r-local model, that considers local permutations, and, in contrast to existing work, does not assume the measurements to be partially shuffled. We propose a novel algorithm based on insights from graph matching and recent advances in alignment via the quadratic assignment problem. Numerical experiments on synthetic data show that the proposed algorithm is robust to additive noise and exploits multiple views for improved recovery. As part of future work, we will explore the theoretical performance bounds for the $r$-local model.

\bibliographystyle{unsrt}
\bibliography{strings.bib}
\end{document}